\documentclass[a4paper,11pt]{article}
\usepackage{jheppub} % for details on the use of the package, please
\usepackage{subfigure}
\usepackage[T1]{fontenc} % if needed

\title{\boldmath Localization of scalar field on the brane-world by coupling with gravity }

\author[a,1]{Heng Guo,\note{Corresponding author.}}
\author[a]{Yong-Tao Lu,}
\author[a]{Cai-Ling Wang,}
\author[a]{and Yue Sun. }
\affiliation[a]{School of Physics,
              Xidian University,
              Xi'an 710071, People's Republic of China}
\emailAdd{hguo@xidian.edu.cn}
\emailAdd{luyt@stu.xidian.edu.cn}
\emailAdd{wcl@stu.xidian.edu.cn}
\emailAdd{syue@stu.xidian.edu.cn}

\abstract{In this paper, we consider a coupling mechanism between the kinetic term and the gravity,
in which a coupling function $F(R)$ is introduced into the kinetic term of the five-dimensional scalar field.
Based on this scenario, we investigate the localization of scalar fields in three specific braneworld models:
the Minkowski brane, the de Sitter brane, and the Anti-de Sitter brane. The brane models considered
here are regular with no singularity for scalar curvature.  For the Minkowshi brane case, the zero
mode can always be localized on the brane, and the massive modes can be localized or quasi-localized on the
brane. For the dS$_4$ brane case, two forms of factor $F(R)$ is considered. The zero mode can always be localized, 
and the massive modes could be quasi-localized on the brane. Besides, with the second coupling factor,
the scalar zero mode could be localized on the both sides of the origin of extra dimension, while the massive modes
could be quasi-localized on the origin. Lastly, for the AdS$_4$ brane case, the
localization of the scalar zero mode requires the consideration of a coupling potential $V(\Phi,\varphi)$,
while the massive modes can still be localized on the brane with an infinite number.

}

\begin{document}
\maketitle
\flushbottom

\section{Introduction}

The idea that our observed four-dimensional (4D) universe
might be a 3-brane, embedded in a higher dimensional
spacetime (the bulk), provides new insights for solving the
gauge hierarchy and cosmological constant problems
\cite{RubakovPLB1983,VisserPLB1985,AkamaLNP1983,Randall1999,
LykkenJHEP2000,AntoniadisPLB1990}. Early theories involving
extra dimensions, namely, the Kaluza$-$Klein (KK) type
theories, were proposed to unify Einstein gravity and
electromagnetism \cite{Appelquist}. In braneworld scenarios,
depending on the specific model, the size of the extra
dimensions can vary. It has been suggested that the radii
of the extra dimensions can be as large as a few Tev$^{-1}$
\cite{AntoniadisPLB1990,AntoniadisPRL2012,YangPRD2012}, or
several millimeters \cite{AntoniadisPLB1998}, or even be
infinitely large \cite{Randall1999}.

In the development of the braneworld theory, two types of
brane models have been proposed: thin brane models and thick
brane models. The well-known thin brane models, namely the
Arkani-Hamed-Dimopoulos-Dvali (ADD) brane model
\cite{ArkaniPLB1998,AntoniadisPLB1998} and the Randall-Sundrum
(RS) brane model \cite{Randall1999}, were primarily developed
to address the hierarchy problem. However, the thickness
of the brane was ignored in these theories, and thin brane
was merely a mathematical idea. In the most fundamental
theory, there seems to exist a minimum scale of length,
thus the thickness of a brane should be considered in more
realistic field models. For this reason, more natural thick
brane models have been suggested \cite{De_Wolfe_PRD_2000,GremmPLBPRD,
paper2002,Csaki_NPB_2000,PRD_koyama,Wang_PRD,ThickBrane,Bazeia,0910.0363,
liu_0911.0269,Liu0907.1952,zhong_fRBrane,1009.1684,SlatyerJHEP2007}.

In braneworld scenarios, the important issue is the localization
of various bulk fields for the purpose of recovering the effective
4D gravity \cite{Kehagias0010112,liu_0911.0269,Liu0907.1952,zhong_fRBrane,
LiuPRD1101.4145,ZhongEPJC2016,Zhong1711.09413,German1210.0721} and building
up the Standard Model \cite{Kehagias0010112,LiuZJHEP2008,
JonesMPRD2013,Pomarol9911294,BajcPLB2000,OdaPLB2000,
SousaSPLB2013,RubinEPJC2015,GherghettaPNPB2000,YoumNPB2000,
AbeKPRD2003,GhorokuY2003,OdaPLB2003,Ghoroku0106145,Liu0708.0065,
Jardim1411.6962,Zhao1712.09843,Fu1207.3152,FuCEPRD064007,FuCEPLB180,
FuCEJHEP021}.
%In a braneworld model, except for the localized zero
%modes of various fields, there are massive KK modes of these fields
%which can move freely along the extra dimension, be localized, or
%be quasi-localized on the brane \cite{LiuPRD0904}. These quasi-localized
%modes are referred to as resonance KK modes. Extensive research has
%been conducted on the resonances of various fields in the context of
%extra-dimensional theories \cite{Almeida2009,CruzPLB2014,XuEPJC2015,
%CsakiPRL2000,ZhangDuPRD2016,ZhongZEPJC2018,GuoZPLB2020,SuiGPRD2020,
%TanGEPJC2021,ChenGEPJC2021,TanEPJC2023}, as well as on the study of these
%massive dynamic resonances around black holes \cite{BarrancoPRD2012,
%BarrancoPRL2012,BarrancoPRD2014,Zhou1308.2863,Gossel1308.6426,
%Sampaio1406.3536,Degollado1408.2589,Sanchis1412.8304,Haung1708.0476,
%Sporea1905.05086}.
Among these investigations, the
localization of scalar fields plays a vital role. It is critical
not only for phenomenological model-building but also for the
dynamic description of the models. Scalar particles or scalar
fields, such as the Higgs boson \cite{HiggsPL1964,HiggsPRL1964,
ATLAS1207.7214}, the only elementary particle in the Standard
Model, scalar perturbations within the framework of gravity
localization in braneworld theories
\cite{liu_0911.0269,LiuPRD1101.4145,ZhongEPJC2016,Zhong1711.09413},
and scalar fields derived from the duality relation in 4D
spacetime \cite{Mukhopadhyaya,Quevedo1011.1491}, constitute
essential and highly desirable subjects. For the reasons stated
above, the investigation into the localization of scalar fields
holds fundamental importance within the realm of physics.

Regarding the localization of the scalar field, a focus point of our discussion, the scalar zero mode can be
localized on brane of different types \cite{BajcPLB2000}. {Specifically, in the case of the Minkowski (M$_4$)
brane, many works \cite{LiangPLB2009,Guo1912.01396} have reported the existence of volcanic-like
effective potential for the scalar KK modes, which leads to the localized zero mode. Moreover, the localization
of scalar zero mode can be realized in six-dimensional brane models \cite{OdaPLB2000,AntoninoPRD2009}, and
in certain case, an additional localized massive mode could exist on the Weyl thick brane \cite{LiuZJHEP2008}.}
In the case of the de Sitter (dS$_4$) brane, the effective potentials belong to the P\"{o}schl-Teller-like
effective potentials \cite{LiangDPLB2009,LiuJCAP20090203,GuoPRD201305}. These potentials lead to the
localization of scalar zero modes on the brane as well as mass gaps in the mass spectra. In the case of the
Anti-de Sitter (AdS$_4$) brane, there exist infinitely deep potential wells \cite{LiuJHEP201002,LiuPRD1101.4145},
capable of trapping an infinite number of massive scalar modes. However, localizing the scalar zero mode on
the AdS$_4$ brane requires the fine-tuning of parameters after introducing a coupling potential \cite{LiuJHEP201002}.
Recently, the authors in \cite{FuCEJHEP021,Chen1912.03859,FuCE2002.03606} introduced the discussion on the
braneworld with codimension-$d$, which could bring about emerging images about the scalar as well as vector
KK modes.

In this paper, we consider three coupling mechanisms: the coupling between the kinetic term and the spacetime, 
the coupling between the five-dimensional (5D) scalar field and the spacetime, and the coupling between the 
5D scalar field and the background scalar field which generates the thick brane. The introducing of the 
latter two coupling mechanisms leads to the localization of scalar zero mode on the AdS$_4$ brane. Then,
we focus on the first one. This coupling introduces a factor $F(R)$ into the kinetic term of the action for
the scalar fields. Based on this scenario, we explore the localization of the scalar field with braneworld
models characterized by three different types of brane geometries: Minkowski, dS$_4$, and AdS$_4$.

{In the case of the Minkowski brane, the scalar zero mode can always be localized on the brane. For the massive modes,
if the coupling parameter $t$ within the factor $F(R)$ is less than $\frac16a^2$. The massive modes can only be
quasi-localized on the brane. Then, for $t=\frac16a^2$, finite number of massive modes can be localized on the
brane. Lastly, for $t>\frac16a^2$, all the massive modes are localized. Besides, if $t<0$,
the effective potential could be positive at the center of the brane, and the localized zero mode is suppressed
at the same position.

In the scenario involving the dS$_4$ brane, we consider two forms for the coupling factor $F(R)$. The effective potential
always be a P\"{o}schl-Teller-like potential. The scalar zero mode remains localizable, and the massive
modes can exist as resonant modes. Especially, when employing the second coupling function, the zero mode is localized
on both sides of the origin of extra dimension, while the massive modes can be quasi-localized on the origin. The
scalar zero mode and the massive KK modes are localized or quasi-localized on different positions of the extra
dimension.

In the case of the AdS$_4$ brane, an asymptotically flat brane model is considered. The coupling factor $F(R)$ tends
to value $1$ and has no effect at the boundaries of extra dimension. The massive modes can always be localized on
the brane, while the localization of the scalar zero mode need the coupling potential $V(\Phi,\varphi)$ to be
introduced. }

The organization of this paper is as follows: We provide a
review of our method in Sec. \ref{LocGF}. The localization
of the scalar zero mode and massive modes is discussed in
Sec. \ref{sec3} and Sec \ref{sec4}, respectively. Applying
our method to the concrete braneworld models are shown in
the next section, in which we discuss the case of Minkowski
brane in Sec. \ref{sec5}, dS$_4$ brane in Sec. \ref{sec6}
and AdS$_4$ brane in Sec. \ref{sec7}. The conclusion is
given in Sec. \ref{Cons}.

%%%%%%%%%%%%%%%%%%%%%%%%%%%%%%%%%%%%%%%%%%%%%%%%%%%%%

\section{The {  Method} }\label{LocGF}

%In general, the geometry of a four-dimensional brane with maximal symmetry
%can be \uline{supposed as M$_{4}$, dS$_{4}$ or AdS$_{4}$.} Here, we assume
%the line element of the five-dimensional (5D) spacetime as

We start with the 5D line element (the $[-,+,+,+,+]$
signature will be assumed below)
\begin{eqnarray}
d s^2={g}_{MN}d x^Mdx^N=
 e^{2 A(y)}\hat{g}_{\mu\nu}d x^{\mu}d x^{\nu}+d y^2,
\label{metric}
\end{eqnarray}
where $\text{e}^{2A(y)}$ is the warp factor, $\hat{g}_{\mu\nu}$
is the induced metric of the 3-brane, and $y=x^5$ denotes
the extra dimension. Throughout this paper, capital Latin
letters $M,N,...=0,1,2,3,5$ and Greek letters $\mu,\nu,...
=0,1,2,3$ are used to represent the bulk and brane indices,
respectively.

With the above metric (\ref{metric}), the 5D scalar curvature $R$
of the bulk can be expressed as
\begin{eqnarray} \label{Ry}
R(y)=4\text{e}^{-2A(y)}\Lambda_4 -8\partial_{y,y}A(y)- 20(\partial_{y}A(y))^{2},
\end{eqnarray}
where $\Lambda_4$ is some constant such that 4D scalar curvature $\hat{R}_4=4\Lambda_4$. In the case of an AdS$_4$ 
brane cosmology, $\Lambda_4$ is negative, while for a dS$_4$ brane cosmology, it is positive. And a Minkowski 
spacetime corresponds to the case $\Lambda_4=0$. The braneworld which holds a 5D spacetime of asymptotically 
constant scalar curvature at infinity suggests that the scalar curvature $R$ satisfies
\begin{equation} \label{Cond1}
 R(y\rightarrow\infty)\rightarrow \left\{
\begin{array}{llcll}
{\mathrm{C}_{R}}_{1}         &  & \mathrm{asymptotically\ dS_{5}~ spacetime}   \\
0                            &  & \text{asymptotically\ flat spacetime}             \\
-{\mathrm{C}_{R}}_{2}                &  & \mathrm{asymptotically\ AdS_{5}~ spacetime},
\end{array} \right.
\end{equation}
where ${\mathrm{C}_{R}}_{1}$ and ${\mathrm{C}_{R}}_{2}$ are positive constants. The brane model considered here has no singularity for
the scalar curvature $R$, which is regular.

%In the introduction, it was mentioned that the localization
%of the scalar zero mode on the AdS$_4$ brane could be realized
%by introducing a coupling potential into the 5D action for
%scalars \cite{LiuJHEP201002}. This fact serves as the starting
%point for our discussions. Furthermore, a non-minimally coupling
%term of the gravity and scalar also is taken into consideration
%for the purpose of possible effects on the localization of
%scalar.
We assume the 5D action for a real scalar field $\Phi$ as
\begin{eqnarray}\label{action_scalar00}
S=\int d^{5}x\sqrt{-g}~
          \bigg[-\frac{1}{2}F(R)g^{MN}\partial_{M}\Phi\partial_{N}\Phi-\xi R\Phi^2-V(\Phi,\varphi)\bigg],
\end{eqnarray}
where the factor $F(R)$ is a coupling function representing the coupling between the kinetic term and the
spacetime. {There have been similar coupling with scalar curvature in Horndeski theory \cite{HorndeskiIJTP10363},
and we suggest a simpler form in this work.} The term $\xi R\Phi^2$ describes the coupling between the 5D 
scalar field $\Phi$ and the spacetime, with $\xi$ the coupling parameter. The potential $V(\Phi,\varphi)$ 
is a coupling potential of 5D scalar field $\Phi$ to itself and to the background scalar field $\varphi$ 
which generates the brane. Besides, the potential $V(\Phi,\varphi)$ should include terms 
$\Phi,\Phi^{2},\Phi^{3},\Phi^{4}$ and $(\varphi\Phi)^{2}$, among which the terms $\Phi$
and $\Phi^{3}$ can be eliminated by considering a discrete symmetry. Thus, we can set \cite{DaviesPRD07}
\begin{eqnarray}\label{coupling potential}
V(\Phi,\varphi)=(\lambda \varphi^{2}-u^{2}) \Phi^{2}+\tau \Phi^{4},
\end{eqnarray}
where $\lambda,u$ and $\tau$ are parameters. To start with, we discuss briefly the effects of the potential
$V(\Phi,\varphi)$ and the coupling term $\xi R\Phi^2$. For simplicity, we set function $F(R)=1$. Then, in order
to obtain the Schr\"{o}dinger-like equation for scalar KK modes, we will follow Ref. \cite{Randall1999} and
perform the following coordinate transformation
\begin{equation} \label{transformation}
\left\{
   \begin{array}{ll}
     & dz=e^{-A(y)}dy \\
     & z=\int e^{-A(y)}dy
   \end{array}
 \right.
\end{equation}
with the boundary condition $z(y=0)=0$, where the coordinate $z$ is a conformal coordinate of $y$. Next,
based on the action (\ref{action_scalar00}), we carry out the similar calculations to that shown in
Ref. \cite{LiuJHEP201002}. The effective potential for KK modes can be expressed as
\begin{eqnarray}\label{Vz-coupling}
V(z)=\frac{3}{2}\partial_{z,z}A(z)+\frac{9}{4}(\partial_{z}A(z))^{2}
    +2e^{2A(z)}(\xi R+\lambda \varphi^{2}-u^{2}).
\end{eqnarray}
Meanwhile, the scalar curvature $R$ in terms of coordinate
$z$ can be obtained:
\begin{eqnarray} \label{Rz}
R(z)=4\text{e}^{-2A(z)}\left(\Lambda_4 -2\partial_{z,z}A(z)- 3(\partial_{z}A(z))^{2}\right).
\end{eqnarray}
And then, the potential (\ref{Vz-coupling}) is recast to
\begin{eqnarray}\label{Vz-coupling1}
V(z)=(\frac{3}{2}-16\xi)\partial_{z,z}A(z)+(\frac{9}{4}-24\xi)(\partial_{z}A(z))^{2}
    +2e^{2A(z)}(\lambda \varphi^{2}-u^{2})+8\xi\Lambda_4.
\end{eqnarray}
Compared to the effective potential presented in Ref.
\cite{LiuJHEP201002}, the effective potential $V(z)$
(\ref{Vz-coupling1}) contains similar terms, but the
coefficients of these terms have been modified due to
the presence of the coupling term $\xi R\Phi^{2}$. Thus,
referring to the results in Ref. \cite{LiuJHEP201002},
we can draw the conclusion that by introducing the
potential $V(\Phi,\varphi)$ and the term $\xi R\Phi^{2}$,
the limits of the effective potential $V(z)$ can be
altered when far away from the brane, while the types
of potential remain unchanged. Consequently, additional
massive scalar modes can be localized on the Minkowski
or dS$_4$ brane, or the scalar zero mode can be localized
on the AdS$_4$ brane through the fine-tuning of parameters,
which holds significant implications.

%there exists a critical value $\lambda_0$ for the parameter
%$\lambda$ that determines the shapes of the potentials of
%the scalar KK modes. For $\lambda>\lambda_0$,  the potentials
%have a negative value at the location of the brane and
%tend to be infinite when far away from the brane. This suggests
%that the mass spectra of the scalars consist of infinite
%discrete bound KK modes. In this case, the scalar zero mode
%can be localized through fine-tuning of parameters. On the
%other hand, for $\lambda<\lambda_0$, the potentials of the
%scalar KK modes have no well, which leads to no localized
%scalar KK modes on the brane.
%\uline{the shapes of effective potential
%could be shifted to form a finite deep negative well at the
%location of thick brane}, and the scalar zero mode could be trapped
%on the brane by fine-tuning of parameters.

Then, we turn attention to the coupling mechanism
between the kinetic term and the spacetime.
In order to describe its effects distinctly, we consider
neither the potential $V(\Phi,\varphi)$ nor the coupling terms
$\xi R\Phi^{2}$ with the parameters $\lambda=0,u=0$, and
$\xi=0$ below. Hence, the 5D action (\ref{action_scalar00})
turns into the following one:
\begin{eqnarray}\label{action_scalar}
S=-\frac{1}{2}\int d^{5}x\sqrt{-g}~
          F(R)g^{MN}\partial_{M}\Phi\partial_{N}\Phi.
\end{eqnarray}

The rules that confirm the form of function $F(R)$ are:
\begin{enumerate}
\item {
      The 5D scalar curvature $R$ and the coupling factor $F(R)$ should be nonsingular.}
\item {In a region without gravity, the 5D spacetime becomes flat and its scalar curvature $R\rightarrow0$, so the
       coupling turns into the minimal coupling with $F(R)\rightarrow1$, and the action (\ref{action_scalar}) returns
       to the form of a free scalar field:}
\begin{equation} \label{action0}
S=-\frac{1}{2}\int d^{5}x\sqrt{-g}~
          g^{MN}\partial_{M}\Phi\partial_{N}\Phi.
\end{equation}
%\item the action should preserve the localization of zero mode of vector fields,
\item { The function $F(R)$ should satisfy the positivity condition
%the integration
%\begin{equation}
%\int_{-\infty}^{+\infty}  F(R)\, \dd y > 0 \label{positivity}
%\end{equation}
\begin{equation}
  F(R) >0 \label{positivity}
\end{equation}
%and the finity condition at the same time }
%\begin{equation} \label{finity}
%\int_{-\infty}^{+\infty}  F(R)\, \dd y  < \infty
%\end{equation}
 to preserve the canonical form of 4D action}.
\end{enumerate}

Note that the restriction mentioned in the third item is less strict in comparison to the second described in Ref.
\cite{Zhao1712.09843}. This is because the metric $g_{MN}$ and the 5D scalar $\Phi$ incorporate extra-dimensional
components. {In Ref. \cite{Zhao1712.09843}, three forms for the function $F(R)$ were proposed by the
authors. Two of these forms belong to polynomials, while the third is an exponential function. However, it is
noteworthy that the exponential function can always be expanded into a polynomial. Additionally, the dilaton scalar
field $\pi$, which shares the same even-parity as the scalar curvature, emerges in the exponential form
$e^{\xi\pi}$ when coupled with the kinetic term \cite{Kehagias0010112,FuCE1101.0336}. Considering this analogy,
the scalar curvature behaves similarly to this dilaton scalar field. Consequently, we will consider $F(R)$ in the
forms of both polynomials and exponential functions. }

%%%%%%%%%%%%%%%%%%%%%%%%%%%%%%%%%%%%%%%%%%%

\subsection{Localization of the Zero Mode}\label{sec3}

In order not to contradict with the present observations, the
zero mode of the scalar field should be confined on the brane,
which will impose some constraints on the function $F(R)$. We
have implicitly assumed that bulk scalar fields considered in
this paper make little contribution to the bulk energy, so
that the solutions given below remain valid even in the
presence of bulk scalar fields.

Let us introduce the decomposition
\begin{equation} \label{zeroDec}
\Phi(x^{\mu},y)=\sum_n \phi_n(x^{\mu})\chi_n(y),
\end{equation}
where $\phi_n(x^{\mu})$ are the 4D scalar fields, and the
scalar KK modes $\chi_n(y)$ are supposed to be functions of
the coordinate $y$. Then, the 5D action
(\ref{action_scalar}) can be reduced to
\begin{equation} \label{action5}
S=-\frac{1}{2}\int d y F(R)e^{2A}\chi_n^2(y)\int d^4x
  \sqrt{-\hat{g}}\left(\hat{g}^{\mu\nu}\partial_{\mu}\phi_n\partial_{\nu}\phi_n
  + m_n^2 \phi_n^2 \right),
\end{equation}
where $\chi_n(y)$ satisfy the following equation
\begin{equation} \label{eq}
\chi_n''+\left(4A'+\frac{F'(R)}{F(R)}\right)\chi_n'
   =-m_n^2\chi_n e^{-2A},
\end{equation}
with the boundary conditions either the Neumann $\chi_n'(\pm \infty)=0$
or the Dirichlet $\chi_n(\pm \infty)=0$, where the prime stands
for the derivative with respect to $y$ in this subsection.

The localization of the scalar field requires
\begin{equation} \label{int}
\int_{-\infty}^{+\infty} d y F(R)e^{2A} \chi_n^2(y)=1.
\end{equation}

For the zero mode, $m_0=0$, so Eq. (\ref{eq}) reads
\begin{equation} \label{eq0}
\chi_0''+\left(4A'+\frac{F'(R)}{F(R)}\right)\chi_0'=0.
\end{equation}
By setting $\gamma'=4A'+\frac{F'(R)}{F(R)}$,
the above equation (\ref{eq0}) becomes
\begin{equation}
\chi_0''+\gamma'\chi_0'=0,
\end{equation}
from which the general solution of zero mode is
\begin{equation} \label{appzeroso1}
\chi_{0}=\mathrm{c}_0 +\mathrm{c}_1 \int e^{-\gamma} dy.
\end{equation}
Here $\mathrm{c}_0$ and $\mathrm{c}_1$ are arbitrary integration
constants. The braneworld discussed here holds $\mathbb{Z}_2$
symmetry along the extra dimension. The functions $F(R),A(y)$
and $\gamma$ are all even functions of the coordinate $y$,
which makes the second term in Eq. (\ref{appzeroso1})
odd. When imposing the Dirichlet boundary condition
$\chi_{n}(\pm\infty)=0$, it leads to $\mathrm{c}_0=0$ and
$\mathrm{c}_1=0$. On the other hand, the Neumann boundary
condition $\chi'_{n}(\pm\infty)=0$ only leads to $\mathrm{c}_1=0$.
Consequently, the zero mode solution is
\begin{equation}
\chi_0=\mathrm{c}_0.
\end{equation}
The localization of zero mode requires the following condition
\begin{eqnarray} \label{int2}
& &\int_{-\infty}^{+\infty} d y F(R)e^{2A}\chi_0^2(y)\nonumber\\
&=&\mathrm{c}_0 ^{2}\int_{-\infty}^{+\infty}  d y F(R) e^{2A}=1 .
\end{eqnarray}
Therefore, whether the above integration (\ref{int2}) converges
is determined by the asymptotic behaviors of both the function
$F(R)$ and the warp factor $e^{2A}$ when far away from the
brane. Specifically, the convergent condition is
\begin{equation} \label{Cd2}
F(R(y\to \pm \infty)) \propto y^{-p}e^{-2 A}
\end{equation}
with $p>1$. {At this point, the normalization condition (\ref{int2}) is equivalent to the following condition
\begin{equation} \label{Cd3}
  \int^{+\infty}_1y^{-p}dy=\frac{1}{p-1}<\infty.
\end{equation}
It can be seen that this condition is met.} Furthermore, from Eq. (\ref{Cd2}), we know that the localization of scalar
zero mode are model$-$dependant. The further discussions will be presented explicitly based on specific braneworld
models in Sec. \ref{Loc}.

%%%%%%%%%%%%%%%%%%%%%%%%%%%%%%%%%%%%%%%%%%%%%

\subsection{Localization of Massive Modes} \label{sec4}

The localization of massive modes is determined by the shapes of the effective potentials for scalar KK modes
in the corresponding Schr\"{o}dinger-like equation.

According to the action (\ref{action_scalar}), by employing
the KK decomposition
\begin{eqnarray} \label{Decom-y}
\Phi(x^\mu,y)=\sum_n \phi_n(x^\mu)\tilde{\chi}_n (y)e^{-2 A(y)} (F(R))^{-\frac{1}{2}}
\end{eqnarray}
and demanding $\phi_n(x^\mu)$ satisfy the 4D massive
Klein$-$Gordon equation
\begin{eqnarray} \label{KGequation}
\bigg(\frac{1}{\sqrt{-\hat{g}}} \partial_{\mu}(\sqrt{-\hat{g}} \hat{g}^{\mu\nu} \partial_{\nu})-m_{n}^{2}\bigg)\phi_{n}=0,
\end{eqnarray}
we can obtain the equation for the scalar KK modes
$\tilde{\chi}_{n}(y)$ as
\begin{eqnarray} \label{eq8}
\left[-\partial^{2}_y+ V(y)\right]{\tilde{\chi}}_n(y)
   = m_{n}^{2} {\tilde{\chi}}_{n}(y) e^{-2 A(y)},
\end{eqnarray}
in which
\begin{eqnarray} \label{Vy}
  V(y)=2 A''(y)+4 A'^{2}(y)
       +\frac{F''(R)}{2 F(R)}+\frac{2 A'(y) F'(R)}{F(R)}
       -\frac{F'^2(R)}{4 F^2(R)}.
\end{eqnarray}
Here the scalar KK modes $\tilde{\chi}_{n}(y)=\chi_{n}(y)e^{2 A(y)}(F(R))^{\frac{1}{2}}$,
$m_{n}$ is the mass of the scalar KK excitations and
the prime denotes derivative with respect to $y$.

%In view of the expression (\ref{Vy}), Eq. (\ref{eq8})
%can be written as $H\tilde{\chi}_{n}(y)=m_{n}^{2}\tilde{\chi}_{n}(y)$
%\cite{De_Wolfe_PRD_2000,Csaki_NPB_2000}, where the Hamiltonian
%operator is given by $H=Q^{\dag}Q$ with
%$Q=-\frac{\dd}{\dd y}+\Gamma'(y)$ with $\Gamma(y)$ a function
%of coordinate $y$. Thus, Eq. (\ref{eq8}) can be recast
%to
Furthermore, Eq. (\ref{eq8}) can be factorized into the
following one
\begin{equation} \label{eq9}
\bigg(\frac{d}{d y}+\Gamma'(y)\bigg)\bigg(-\frac{d}{d y}
  +\Gamma'(y)\bigg)\tilde{\chi}_n(y)
 = m_n^2 \tilde{\chi}_n(y) e^{-2 A(y)},
\end{equation}
where there is
\begin{equation}
\Gamma'(y)=2 A'(y)+\frac{F'(R)}{2 F(R)}.
\end{equation}
Without loss of the generality, we can take
\begin{equation}
\Gamma(y)=2 A(y) + \frac{1}{2}\ln{F(R)}.
\end{equation}
Then, the solution of the scalar zero mode can be obtained
from Eq. (\ref{eq9}) as
\begin{equation} \label{zeroy1}
\tilde{\chi}_0(y)=N_1 e^{2 A(y)} (F(R))^{\frac{1}{2}},
\end{equation}
where $N_1$ is the normalization constant. The localization
of scalar zero mode $\tilde{\chi}_0(y)$ requires
\begin{equation}
\int d y \tilde{\chi}^2_{0}(y)= N_1^{2}\int d y F(R)e^{4A(y)} =1.
\end{equation}

On the other hand, it can be observed that the potential $V(y)$
in Eq. (\ref{eq8}) represents the effective potential for the
scalar zero mode, rather than for the massive ones. To address
this issue, we will proceed with our discussions in terms of
the conformal coordinate $z$, instead of the coordinate $y$.
By performing the coordinate transformation (\ref{transformation}),
the metric (\ref{metric}) can be expressed as
\begin{equation} \label{gcf}
d s^2=e^{2 A(z)}(\hat{g}_{\mu\nu}d x^{\mu}d x^{\nu}+d z^2).
\end{equation}
And we can obtain the equation of motion derived from the action
(\ref{action_scalar}) as
\begin{equation} \label{EquMo1}
\frac{1}{\sqrt{-\hat{g}}} \partial_{\mu}(\sqrt{-\hat{g}} \hat{g}^{\mu\nu} \partial_{\nu}\Phi)
+\frac{e^{-3A(z)}}{F(R)} \partial_{z}(e^{3A(z)}\partial_{z}\Phi) =0.
\end{equation}
%Based on the decomposition (\ref{zeroDec}), we choose
%$\tilde{\chi}_{n}(z)=\chi_{n}(y)e^{\frac{3}{2} A(z)} (F(R))^{\frac{1}{2}}$.
Through the KK decomposition
\begin{equation} \label{MassDecz}
\Phi(x^\mu,z)=\sum_n \phi_n(x^\mu){\tilde{\chi}}_n(z)e^{-\frac{3}{2} A(z)} (F(R))^{-\frac{1}{2}},
\end{equation}
and demanding $\phi_n(x^\mu)$ satisfies the 4D Klein-Gordon
equation (\ref{KGequation}), we can further obtain the equation
for scalar KK modes $\tilde{\chi}_{n}(z)$:
\begin{eqnarray} \label{eq3}
\left[-\partial^{2}_z+ V(z)\right]{\tilde{\chi}}_n(z)
  =m_{n}^{2} {\tilde{\chi}}_{n}(z),
\end{eqnarray}
which is a Schr\"{o}dinger-like equation with the effective
potential given by
\begin{eqnarray} \label{Vz}
  V(z)=\frac{3}{2} A''(z)+\frac{9}{4}A'^{2}(z)
       +\frac{F''(R)}{2 F(R)}+\frac{3 A'(z) F'(R)}{2 F(R)}
       -\frac{F'^2(R)}{4 F^2(R)},
\end{eqnarray}
where the prime denotes the derivative with respect to $z$ here.

Similarly, the Schr\"{o}dinger-like equation (\ref{eq3}) can be recast to
\begin{equation} \label{eq4}
\bigg(\frac{d}{d z}+\Gamma'(z)\bigg)\bigg(-\frac{d}{d z}
 +\Gamma'(z)\bigg)\tilde{\chi}_n (z)= m_n^2\tilde{\chi}_n (z),
\end{equation}
where there is
\begin{equation} \label{Gammaz}
\Gamma'(z)=\frac{3}{2}A'(z)+\frac{F'(R)}{2 F(R)}.
\end{equation}
%As Eq. (\ref{eq4}) shows, the Hamiltonian operator $H=Q^{\dag}Q$
%with $Q=-\frac{\dd}{\dd z}+\Gamma'(z)$ with $\Gamma(z)$ a
%function of coordinate $z$. And we can see that since the
%operator $H$ is positive definite, there are no normalizable
%modes with negative $m^{2}$, namely, there is no tachyonic
%scalar mode. Thus, scalar zero mode is the lowest mode in
%the spectrum.
From the above expression (\ref{eq4}), we can see that
there are no normalizable modes with negative $m_n^{2}$,
namely, there is no tachyonic scalar mode. Thus, the scalar
zero mode is the lowest mode in the spectrum.

From Eq. (\ref{Gammaz}), without loss of
the generality, we can take
\begin{equation}
\Gamma(z)=\frac{3}{2}A(z) + \frac{1}{2}\ln{F(R)}.
\end{equation}
And based on Eq. (\ref{eq4}), the solution of the scalar
zero mode is easy to be found:
\begin{equation} \label{zeroz1}
\tilde{\chi}_0(z)=N_2e^{\frac{3}{2} A(z)} (F(R))^{\frac{1}{2}},
\end{equation}
with $N_2$ the normalization constant. This zero mode solution
can be related to the one expressed in terms of the coordinate
$y$ (\ref{zeroy1}) through the decomposition (\ref{Decom-y})
and (\ref{MassDecz}).

Returning to the discussion of the massive modes, the action
of the 5D scalar field (\ref{action_scalar}) can be reduced
to
\begin{equation} \label{action7}
S=-\frac{1}{2}\int d z \tilde{\chi}_n^2(z)\int d^4x
  \sqrt{-\hat{g}} \left(\hat{g}^{\mu\nu}\partial_{\mu}\phi_n\partial_{\nu}\phi_n
  + m_n^2 \phi_n^2 \right),
\end{equation}
in which the Schr\"{o}dinger-like equation (\ref{eq3}) should
be satisfied. In order to get the effective action for the 4D
scalar field, the orthonormality condition needs to be
introduced:
\begin{equation}
\int d z \tilde{\chi}_m(z) \tilde{\chi}_n(z) = 0. ~~~(m\neq n) %\delta_{mn}.
\end{equation}
And this equation contains the localization condition for $\tilde{\chi}_n(z)$
\begin{equation} \label{LocKK}
\int d z \tilde{\chi}^2_n(z)<\infty.
\end{equation}

The existence of localized massive scalar KK modes is determined
by the shapes of the effective potential $V(z)$ (\ref{Vz}),
which is influenced by both the functions $F(R(z))$ and $A(z)$.
With specific braneworld models and the corresponding functions
$F(R)$, the localization of massive scalar KK modes will be
discussed below.

%%%%%%%%%%%%%%%%%%%%%%%%%%%%%%
\section{Concrete Braneworld Models}\label{Loc}

In this section, we will describe the localization of the scalar
field while considering the coupling between the kinetic
term and the spacetime. These processes will be presented using
specific braneworld models that incorporate three different cases
of brane geometries: Minkowski, dS$_4$ and AdS$_4$, respectively.

{
We start with the following 5D action of thick branes, which are generated by a real scalar field $\varphi$
with a scalar potential $V(\varphi)$,
\begin{equation}
 S_{\text{brane}}= \int d^5 x \sqrt{-g}\left[\frac{1}{2\kappa_5^2} R-\frac{1}{2}
     g^{MN}\partial_M \varphi \partial_N \varphi - V(\varphi) \right],
\label{action}
\end{equation}
where $\kappa_5^2=8
\pi G_5$ with $G_5$ the 5D Newton constant.  For simplicity, this constant is set to $\kappa_5^2 = 1$ and the scalar
field is considered to be a function of the extra dimension only, i.e., $ \varphi =\varphi(y)$. The most general 5D
metric can be taken as (\ref{metric}), and the Einstein field equations reduce to the following coupled nonlinear
differential equations:
\begin{subequations}\label{EinsteinEqy}
\begin{eqnarray}
\label{EinsteinEqy_a}
3A''+6A'^{2}-\text{e}^{-2A}\Lambda_4 &=&-\frac{1}{2}\varphi'^{2}-V(\varphi),\\
\label{EinsteinEqy_b}
6 A'^{2}-2\text{e}^{-2A}\Lambda_4 &=& \frac{1}{2}\varphi'^{2}-V(\varphi),\\
\label{EinsteinEqy_c}
\varphi''+4A'\varphi'&=& \frac{d V}{d \varphi},
\end{eqnarray}
\end{subequations}
where the prime stands for the derivative with respect to the extra dimension coordinate.

The 5D metric can also be transformed to the conformal one (\ref{gcf}) with the coordinate
transformation (\ref{transformation}), and the Einstein field equations in the $z$ coordinate can be written as
\begin{subequations}\label{EinsteinEqz}
\begin{eqnarray}
\label{EinsteinEqz_a}
3A''+3A'^{2}-\Lambda_4 &=&-\frac{1}{2}\varphi'^{2}-\text{e}^{2A}V(\varphi),\\
\label{EinsteinEqz_b}
6 A'^{2}-2\Lambda_4 &=& \frac{1}{2}\varphi'^{2}-\text{e}^{2A}V(\varphi),\\
\label{EinsteinEqz_c}
\text{e}^{-2A}(\varphi''+3A'\varphi')&=& \frac{d V}{d \varphi}.
\end{eqnarray}
\end{subequations}

When we discuss the coupling between the scalar fields and the spacetime, it is necessary to make sure the stability of
the brane model against the fluctuations of the scalar modes of metric, and of the background scalar field. Under small
fluctuations of the scalar modes of the metric, the line element is \cite{PRD_koyama}
\begin{eqnarray}
 ds^{2}=
 \text{e}^{2A}[(1+2\alpha)dz^{2}+(1+2\beta)\hat{g}_{\mu\nu}dx^{\mu}dx^{\nu}].
\end{eqnarray}
Then, as shown in Ref.~\cite{PRD_koyama}, the following corresponding linearized 5D Einstein-scalar equations
can be obtained:
\begin{eqnarray}
 &~& \delta\varphi=\frac{3}{\varphi'}(\alpha A'-\beta'), \quad\quad
 %\\ &~&
 \alpha+2\beta=0, \\
 &~&  \beta'' + \Box\beta
        -\left(3A'+2\frac{\varphi''}{\varphi'}\right)\beta'
   +\left(4A'\frac{\varphi''}{\varphi'}-4A''-2\Lambda_4\right)\beta=0,
 \label{masterEq}
\end{eqnarray}
where $\delta\varphi$ denotes the perturbations of the background scalar field,
$\Box\equiv\hat g^{\mu\nu}\bigtriangledown_{\mu}\bigtriangledown_{\nu}$ and $\bigtriangledown_{\mu}$ represents the
covariant derivative with respect to the 4D metric $\hat g_{\mu\nu}$. To examine the stability
of this system, Eq. (\ref{masterEq}) can be transformed into the following form:
\begin{eqnarray}\label{SchEqDeltaPhi}
 [-\partial_{z}^{2}+V_{\text{eff}}(z)]\omega(x^{\mu},z)
  =\Box \omega(x^{\mu},z),
\end{eqnarray}
where $\omega(x^{\mu},z)$ and $V_{\text{eff}}(z)$ are defined as
\begin{eqnarray}
\label{omega}
 \omega &\equiv&
 \frac{1}{\varphi'(z)}\text{e}^{\frac{3A}{2}}\beta(x^{\mu},z),
   \\
\label{Veff}
 V_{\text{eff}}&\equiv&
  -\frac{5}{2}A''+\frac{9}{4}A'^{2}+A'\frac{\varphi''}{\varphi'}
  -\frac{\varphi'''}{\varphi'}+2\left(\frac{\varphi''}{\varphi'} \right)^{2}
   - 2\Lambda_4. ~~~
\end{eqnarray}
Performing the decomposition $\omega(x^{\mu},z)=X(x^{\mu})\zeta(z)$, we can further derive the 4D equation
and the extra-dimensional one, as follows:
\begin{eqnarray}
  \Box X(x^{\mu}) &=& m^2X(x^{\mu}),   \label{4thEqScaPertMin_a}   \\
  (-\partial^2_z+V_{\text{eff}}(z))\zeta(z) &=& m^2\zeta(z).    \label{4thEqScaPertMin_b}
\end{eqnarray}
The latter (\ref{4thEqScaPertMin_b}) is the Schr\"{o}dinger-like equation. Then, concerning the stability of the brane
models under the scalar perturbations, we will analyze the effective potentials within this equation of
three types of branes separately.

}

%%%%%%%%%%%%%%%%%%%%%%%%%%%%%%%%%%%%%%%%%%
\subsection{Minkowski Brane}\label{sec5}

{
In the case of the constant $\Lambda_4= 0$, the brane is flat
(Minkowski brane) and the 5D line element (\ref{metric})
turns into
\begin{eqnarray} \label{Min-metric}
d s^2=e^{2 A(y)}\hat{\eta}_{\mu\nu}d x^{\mu}d x^{\nu}+d y^2,
\end{eqnarray}
where $\hat{\eta}_{\mu\nu}$ denotes the metric of the Minkowski brane. Based on this metric, the scalar curvature (\ref{Ry}) becomes
\begin{eqnarray} \label{Ry-Min}
R(y)=-8A''(y)- 20A'^{2}(y).
\end{eqnarray}

Here, we consider a trial warp factor of Minkowski brane and
it takes the form \cite{SlatyerJHEP2007}
\begin{eqnarray}
 A(y)&=& -a^{2} \left(\frac{2}{3}\ln(\cosh(b y))
      -\frac{1}{6}( \textrm{sech}(b y))^{2}+c\right),  \label{MineA}  \\
 \varphi(y)&=& \sqrt{3} a\tanh(b y), \label{varphiA}
\end{eqnarray}
where $a,b$ and $c$ are constants, among which $a$ is dimensionless, $b$ has dimensions of inverse length, and
$c$ is a dimensionless integration constant. This warp factor has even-parity with respect to $y$, and  we will
solely concentrate on the asymptotic behaviors of the model as $y\rightarrow +\infty$ in the following discussion.

In this flat brane case, the 4D cosmological constant vanishes. Then, the effective potential
within Eq. (\ref{Veff}) becomes
\begin{eqnarray} \label{VeffScaPertMin}
 V_{\text{eff}}(z)=
  -\frac{5}{2}A''+\frac{9}{4}A'^{2}+A'\frac{\varphi''}{\varphi'}
  -\frac{\varphi'''}{\varphi'}+2\left(\frac{\varphi''}{\varphi'} \right)^{2}.
\end{eqnarray}
As there is no analytical expression for the warp factor (\ref{MineA}) in terms of $z$, we re-write the above potential
$V_{\text{eff}}(z)$ concerning $y$ as
\begin{eqnarray} \label{VeffyScaPertMin}
 V_{\text{eff}}(z(y))&=&
  -\frac{e^{2A(y)}[4\partial_{y,y,y}\varphi+8\partial_yA(y)\partial_{y,y}\varphi
  +\partial_y\varphi(14\partial_{y,y}A(y)+5(\partial_yA(y))^2)]}{4\partial_y\varphi}    \nonumber   \\
  & &+\frac{2e^{4A(y)}(\partial_{y,y}\varphi+\partial_yA(y)\partial_y\varphi)^2}{(\partial_y\varphi)^2}.
\end{eqnarray}
Substituting Eqs. (\ref{MineA}) and (\ref{varphiA}) into the above expression, we can obtain the following asymptotic
behaviors of potential $V_{\text{eff}}(z(y))$ as $y\rightarrow+\infty$:
\begin{eqnarray} \label{AsypExprsVeffyScaPertMin}
 V_{\text{eff}}(z(y\rightarrow+\infty))\sim
   e^{-\frac43a^2by}\rightarrow 0.
\end{eqnarray}
The profile of effective potential $V_{\text{eff}}(z)$ (\ref{VeffyScaPertMin}) is depicted in Fig. \ref{figVeffScaPMin} via
numerical methods with parameters $a=1,b=2$, and $c=1$. It can be seen that the potential $V_{\text{eff}}(z)$ tends to zero
when far away from the brane. In addition, concerning the equation (\ref{4thEqScaPertMin_b}), it can further be recast as
\begin{eqnarray} \label{SchroScaPertMin}
 \bigg(-\partial_z+\frac{J'}{J}\bigg)\bigg(\partial_z+\frac{J'}{J}\bigg)\zeta(z)=m^2\zeta(z),
\end{eqnarray}
where $J=e^{\frac32A}\varphi'/A'$. Therefore, there is no KK modes with negative $m^2$ and the brane is stable against the
scalar perturbations.
\vspace{0.2cm}
\begin{figure} [htbp]
\begin{center}
\includegraphics[width= 0.45\textwidth]{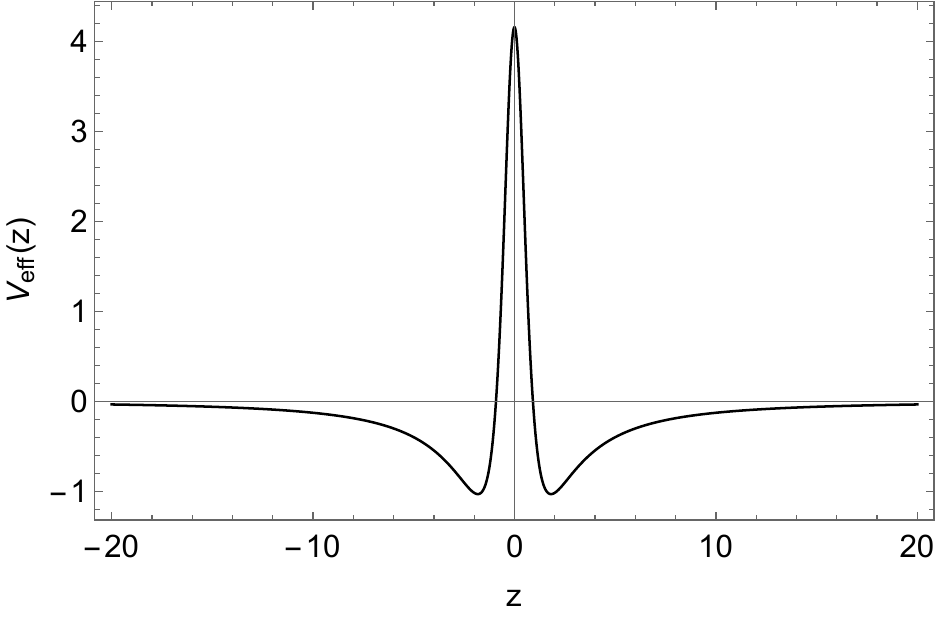}
\caption{{The effective potential $V_{\text{eff}}(z)$ (\ref{VeffyScaPertMin}) corresponding to the
         stability under the scalar perturbations in Minkowski brane case. The parameters
         are set as $a=1,b=2$, and $c=1$.}}
\label{figVeffScaPMin}
\end{center}
\end{figure}
}

\vspace{0.3cm}
With this warp factor (\ref{MineA}), the scalar curvature (\ref{Ry-Min}) becomes
\begin{eqnarray} \label{MinR}
 R(y)=\frac{4}{9}a^{2} b^{2} \bigg( 18( \textrm{sech}(b y))^{4}-5a^{2}\left(2+
   ( \textrm{sech}(b y))^{2}\right)^{2}(\tanh(b y))^{2} \bigg).
\end{eqnarray}
And then, its asymptotic solution as $y\rightarrow+\infty$ is
\begin{eqnarray}\label{ApMinR}
 R(y\rightarrow+\infty)\rightarrow -l+ m e^{-4b y},
\end{eqnarray}
where parameters $l=\frac{80}{9}a^{4}b^{2},m=\frac{64(15a^{2}+18)}{9}a^{2}b^{2}$, for simplicity. {Thus, the 5D
spacetime is asymptotically AdS$_5$ at infinity.}

Meanwhile, the asymptotic expression of $e^{2A(y)}$ as
$y\rightarrow+\infty$ can be given by
\begin{eqnarray}\label{ApMinWarp}
 e^{2A(y)}\rightarrow p e^{-\frac43a^2b y},
\end{eqnarray}
where positive constant $p=e^{-\frac{a^{2}}{3}(6c-4\ln{2})}$.

In this case of Minkowski brane, we suggest the function $F(R)$
takes the form of
\begin{eqnarray}\label{MinFR}
 F(R)=e^{\mathrm{C}_1^t(1- (\frac{l}{R+l})^t)},
\end{eqnarray}
where $t$ is the coupling parameter, and $\mathrm{C}_{1}$ is an arbitrary positive constant. As $y\rightarrow+\infty$, we
can obtain the asymptotic solution of $F(R)$:
\begin{eqnarray}\label{ApMinFR}
 F(R)\rightarrow e^{(\mathrm{C}_{1})^{t}}e^{-n e^{4bt y}},
\end{eqnarray}
where $n=(\frac{5a^2\mathrm{C}_{1}}{12(5a^2+6)})^{t}$. {Thus, as $y\rightarrow+\infty$, the function $F(R)$ converges to limit
zero. We can also see that if the coupling parameter $t=0$, the function $F(R)=1$, and the coupling can be reduced to the minimal coupling. }

Returning to the discussion of the scalar zero mode (\ref{zeroy1}), based on Eqs. (\ref{ApMinWarp}) and (\ref{ApMinFR}),
the asymptotic solution for the scalar zero mode $\tilde{\chi}_{0}(y\rightarrow+\infty)$ can be given by
\begin{eqnarray}\label{ApMinChi}
\tilde{\chi}_{0}(y\rightarrow+\infty)\rightarrow N_1p e^{\frac12\mathrm{C}_{1}^t-\frac{4}{3}a^{2}b y
    -\frac{1}{2}(\frac{5a^2\mathrm{C}_{1}}{12(5a^2+6)})^{t} e^{4bt y}}.
\end{eqnarray}
The localization of the zero mode requires
\begin{eqnarray}\label{LocMinChi}
\int d y \tilde{\chi}_0^2(y)< \infty .
\end{eqnarray}
{From Eq. (\ref{ApMinChi}), there is
\begin{eqnarray}\label{PDChi0Min}
\tilde{\chi}_{0}^2(y\rightarrow+\infty)\rightarrow N_1^2p^2e^{\mathrm{C}_{1}^t-\frac{8}{3}a^{2}b y
    -(\frac{5a^2\mathrm{C}_{1}}{12(5a^2+6)})^{t} e^{4bt y}}.
\end{eqnarray}
{For this expression, as $y\rightarrow+\infty$, for $t<0$ the exponential factor within the r.h.s decays
exponentially, and for $t>0$ it vanishes super exponentially.} Consequently, the scalar zero mode can always be localized
on the thick brane regardless of whether parameter $t$ is positive or negative. We plot
the zero mode in terms of both the coordinate $y$ and the coordinate $z$ in Fig. \ref{figVy1rhoy1}. The parameters used
for the plots are $\mathrm{C}_{1}=2,a=1,b=2,c=1$, and the coupling parameter $t$ is set to $\frac{a^{2}}{10},\frac{a^{2}}{6}$
and $\frac{a^{2}}{4}$, respectively.
\begin{figure} [htbp]
\begin{center}
\includegraphics[width= 0.45\textwidth]{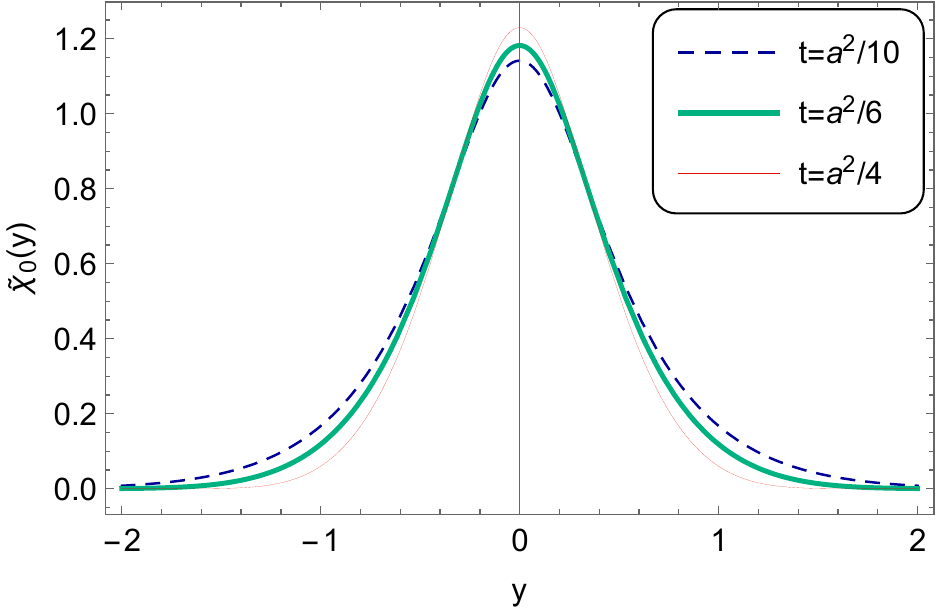}
\includegraphics[width= 0.45\textwidth]{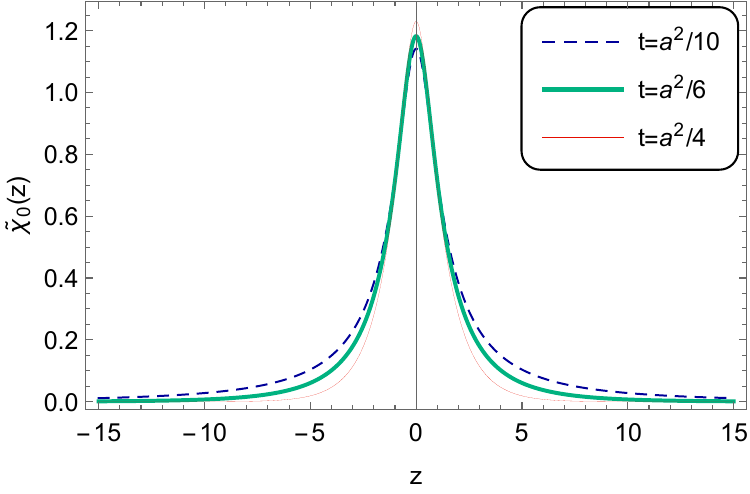}
\caption{Plots of the zero mode in terms of the coordinate $y$
  ($\tilde{\chi}_{0}(y)$ for left figure) and the coordinate $z$
  ($\tilde{\chi}_{0}(z)$ for right figure) with parameters
  $\mathrm{C}_{1}=2, a=1, b=2$ and $c=1$. The coupling
  parameter is set as $t=\frac{a^{2}}{10}$ for the
  dashed line, $t=\frac{a^{2}}{6}$ for the thick
  line and $t=\frac{a^{2}}{4}$ for the thin line,
  respectively.}
\label{figVy1rhoy1}
\end{center}
\end{figure}

However, in the context of the braneworld model considered
here, obtaining an analytic solution for the warp factor with
respect to $z$, as well as for the effective potential $V(z)$
(\ref{Vz}), is challenging. Consequently, we will employ numerical
methods to calculate the solution for the potential $V(z)$.

Carrying out the coordinate transformation (\ref{transformation}),
we can express the effective potential $V(z)$ in terms
of $z$ as follows
\begin{eqnarray}\label{MinVz-y}
 V(z(y))=&&\frac{15}{4}e^{2A(y)} A'^{2}(y)+\frac{3}{2}e^{2A(y)} A''(y) \nonumber\\
         &&+\frac{e^{2A(y)} F''(R)}{2F(R)}+\frac{2e^{2A(y)} A'(y) F'(R)}{F(R)} \nonumber\\
         &&-\frac{e^{2A(y)}F'^{2}(R)}{4F^{2}(R)}.
\end{eqnarray}
Next, by substituting the asymptotic solutions of the warp factor (\ref{ApMinR}) and the function $F(R)$ (\ref{ApMinFR}) into the
potential $V(z(y))$ (\ref{MinVz-y}), we can get its expression. {Furthermore, at $y=0$, the effective potential $V(z(y))$ becomes
\begin{eqnarray}\label{MinVzyt}
 V(z(y=0))= -\frac{3b^2e^{\frac{1}{3}a^2(1-6c)}}{20a^2+18}\bigg(a^2(10a^2+9)+3(5a^2+4)\bigg(\frac{10a^2\mathrm{C}_{1}}{10a^2+9}\bigg)^tt\bigg).
\end{eqnarray}
From this expression, it can be seen that if parameter $t>0$, the effective potential will always be negative at the origin of the
extra dimension. If parameter $t<0$, the effective potential could be positive with certain value of parameters.

Moreover, as $y\rightarrow+\infty$, the asymptotic solution for the effective potential can be obtained:
\begin{eqnarray}\label{ApMinVz-y}
  V(z(y\rightarrow+\infty))\rightarrow 2^{2+\frac{4a^2}{3}-4t}\bigg(\frac{5a^2\mathrm{C}_1}{3(5a^2+6)}\bigg)^{2t}b^2t^2e^{-2a^2c}
        e^{4by(2t-\frac13a^2)}.
\end{eqnarray}
From this expression, it can be concluded that the effective potential has the following
asymptotic behaviors:
\begin{equation} \label{Vzinf2}
 V(z(y\rightarrow\infty))\rightarrow \left\{
\begin{array}{llcll}
+\infty         & &t & > \frac{a^{2}}{6}   \\
\mathrm{C}        & &t &= \frac{a^{2}}{6}  \\
0                 & &t &< \frac{a^{2}}{6},
\end{array} \right.
\end{equation}
with the positive constant $\mathrm{C}$:
\begin{eqnarray}\label{LmtPTMinSca}
 \mathrm{C}=\frac19a^4b^2\bigg(\frac{20a^2\mathrm{C}_1}{3(5a^2+6)}\bigg)^{a^2/3}e^{-2a^2c}.
\end{eqnarray}
In terms of the analysis on Eq. (\ref{MinVzyt}), we will discuss the localization of KK modes in cases of $t>0$, and $t<0$ below. }

For the former scenario where $t>0$, we depict the shapes of both the effective potential $V(z(y))$ and the numerical solution
of the effective potential $V(z)$ in Fig. \ref{figVy-z}. The parameters used are $\mathrm{C}_{1}=2,a=1,b=2,c=1$, and coupling
parameter $t$ is set to $\frac{a^{2}}{10},\frac{a^{2}}{6}$ and $\frac{a^{2}}{4}$, respectively. We can observe that the shapes
of the two expressions for the effective potential are similar. However, the one expressed in terms of the coordinate $z$
extends along the extra dimension in comparison to the other. { As the figures show, there are three types of
effective potential dependent on the parameter $t$. Subsequently, we will investigate the localization of massive KK modes
in these three cases separately.
\begin{figure} [htbp]
\begin{center}
\includegraphics[width= 0.45\textwidth]{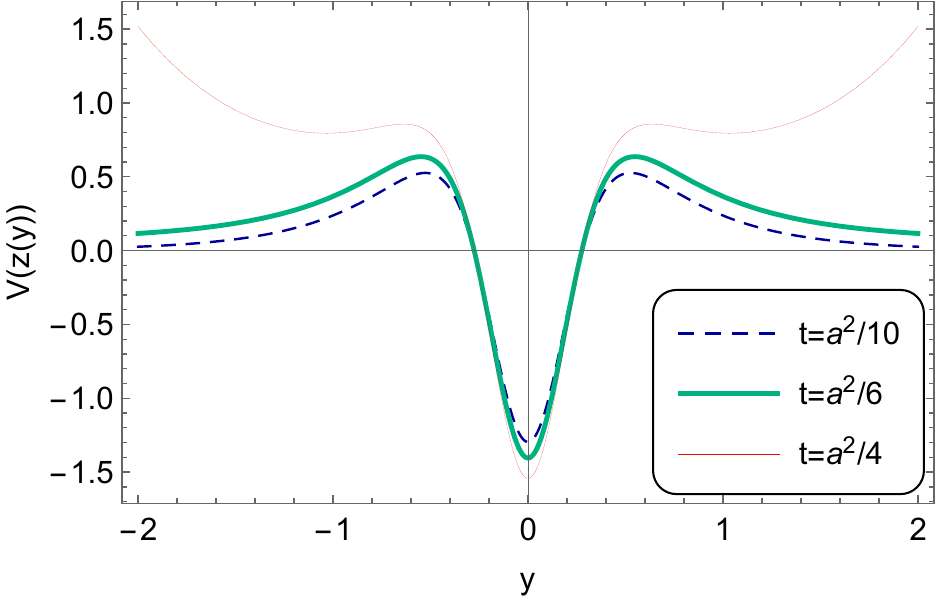}
\includegraphics[width= 0.45\textwidth]{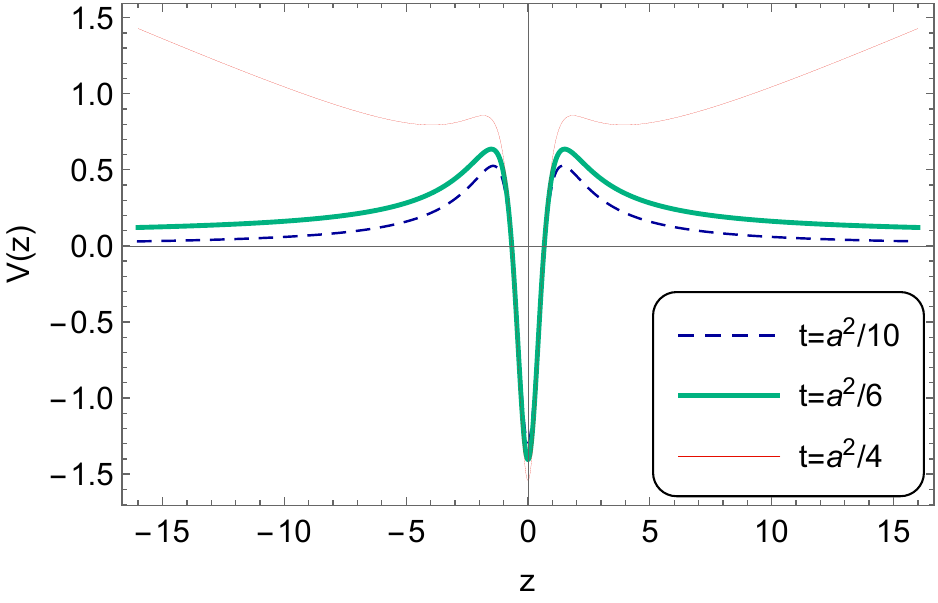}
\caption{Plots of the effective potential $V(z(y))$ and
  the numerical solution of effective potential $V(z)$
  with parameters $\mathrm{C}_{1}=2, a=1, b=2$ and $c=1$.
  The coupling parameter is set as $t=\frac{a^{2}}{10}$
  for the dashed line, $t=\frac{a^{2}}{6}$ for the thick
  line, and $t=\frac{a^{2}}{4}$ for the thin line,
  respectively.}
\label{figVy-z}
\end{center}
\end{figure}

Initially, for the case of $t<a^2/6$, the effective potential takes the volcano shape. There is no localized massive modes on the
thick brane. However, two barriers exist near the origin of the extra dimension. This configuration facilitates the existence of
scalar resonant modes. % which describe massive 4D scalars with finite lifetimes on the brane \cite{TanQEPJCEvolution}.
Following the approach outlined in Ref. \cite{LiuPRD0904,LiuPRD2009.80}, the relative probability function for a scalar resonance on the
thick brane is defined as
\begin{eqnarray} \label{ResonanceP}
P(m^2)=\frac{\int^{z_{b}}_{-z_{b}} |\tilde{\chi}(z)|^{2} dz}{\int^{z_{max}}_{-z_{max}} |\tilde{\chi}(z)|^{2} dz},
\end{eqnarray}
Here, $2z_{b}$ approximately represents the width of the thick brane, and $z_{max}=10z_{b}$. For these KK modes with larger $m^{2}$
than the maximum of the corresponding potential, they will asymptotically turn into plane waves and the probabilities
for them trend to the value of $0.1$. The lifetime $\tau$ of a resonance state is $\tau \sim \Gamma^{-1}$ with $\Gamma=\delta m$
being the full width at half maximum of the resonant peak.

With the method presented above, we can solve for the resonant modes with specific value of parameters. Additionally, for the case
($t<a^2/6$) considered here, the value of partial parameters differ from those in Fig. \ref{figVy-z}, in order for a more obvious
description. The relative possibilities, the mass spectra alongside the effective potential are illustrated in Fig. \ref{ResoVolMin}.
Each peak within the curve of relative possibilities represents a resonant KK mode. In the mass spectra, the ground state is
the scalar zero mode, and the massive modes are resonant KK modes. Figure \ref{ResoVolMin} displays three resonant modes, which are
visually depicted in Fig. \ref{WavFuncResoVolMin}. Detailed information for the mass, width, and lifetime of all resonant modes
is provided in Table \ref{tabResoVolMin}. Therefore, in the case of $t<a^2/6$, the scalar zero mode can be localized on the brane,
and the massive KK modes could exist as resonant modes on the brane.
\begin{figure} [htbp]
\begin{center}
\includegraphics[width= 0.45\textwidth]{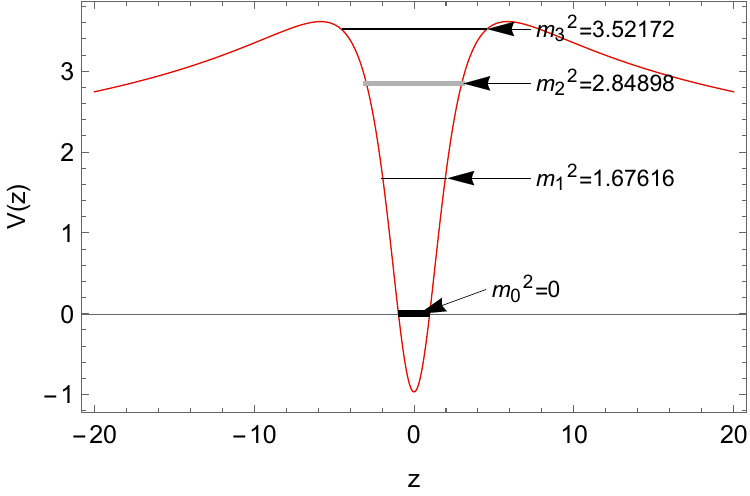}
\includegraphics[width= 0.45\textwidth]{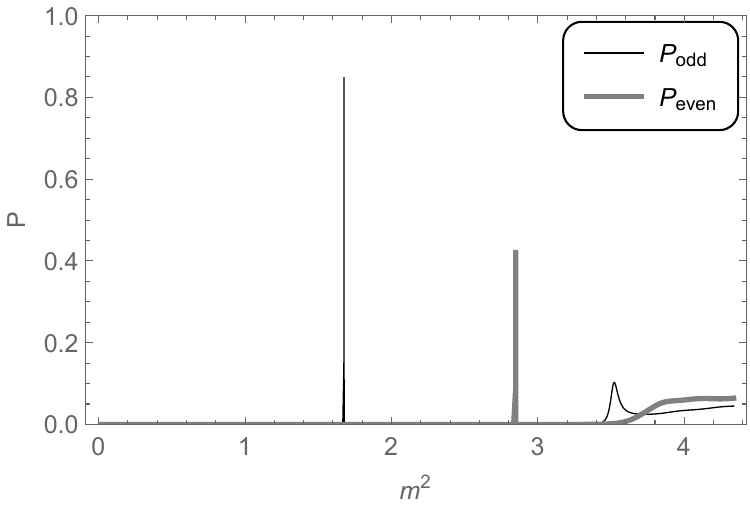}
\caption{The mass spectra, the effective potential $V(z)$, and corresponding relative probability $P$
         with parameters $a=1.6,b=2,c=1,\text{C}_1=3\times10^3,t=a^2/7$. The potential for the red line,
         the zero mode for the thick black line, the odd parity resonant mode for the thin black lines,
         and the even parity resonant mode for the thick grey line.}
\label{ResoVolMin}
\end{center}
\end{figure}

\begin{figure}[htb]
\begin{center}
\includegraphics[width= 0.3\textwidth]{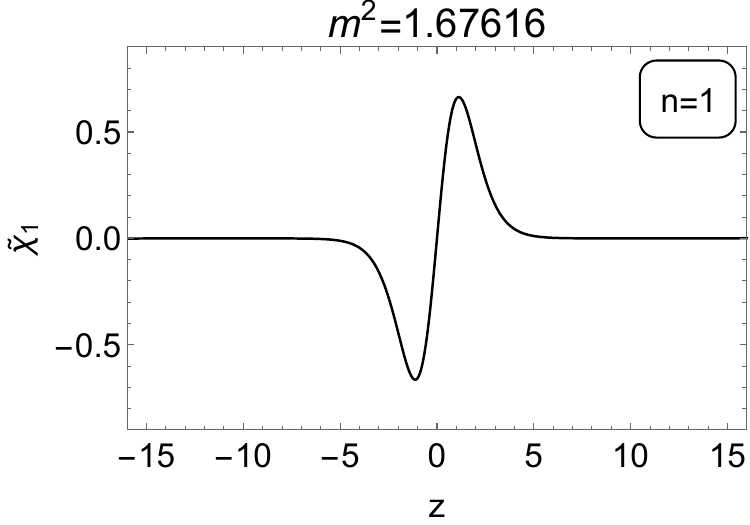}
\includegraphics[width= 0.3\textwidth]{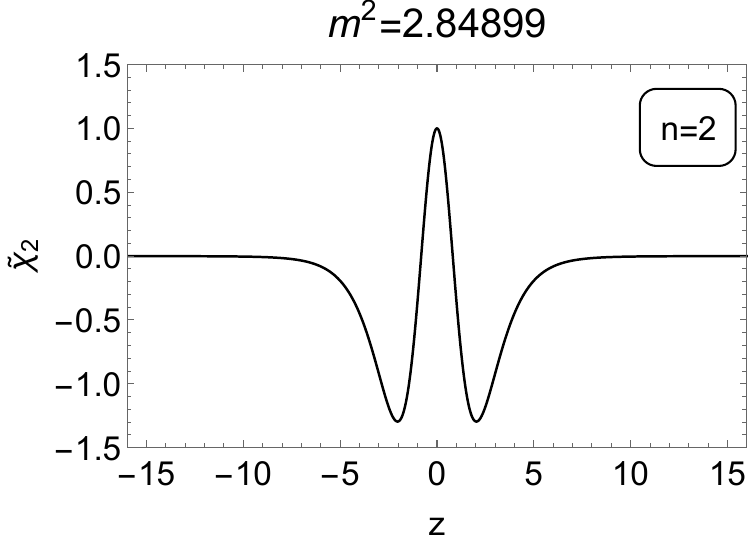}
\includegraphics[width= 0.3\textwidth]{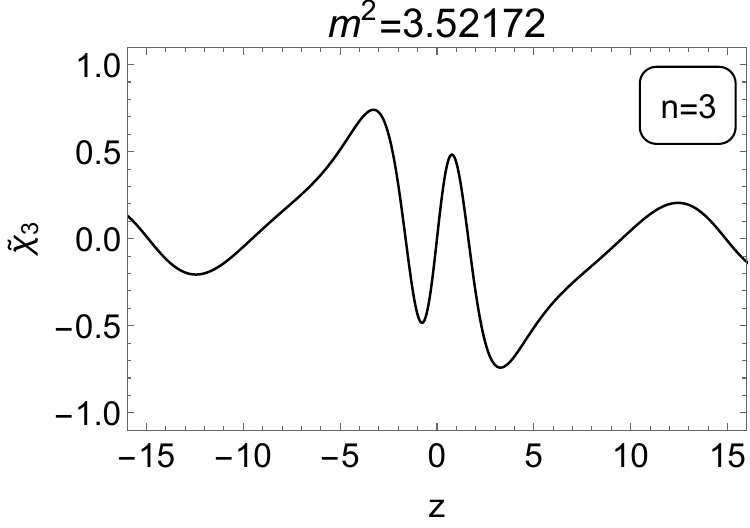}
\end{center}\vskip -2mm
\caption{The shapes of resonant KK modes $\tilde{\chi}(z)$ for scalars
         with different $m^{2}$. The parameters are set as $a=1.6,b=2,c=1,
         \text{C}_1=3\times10^3,t=a^2/7$.}
 \label{WavFuncResoVolMin}
\end{figure}

\begin{table}[ht]
\begin{center}
\caption{The mass, width, and lifetime for scalar resonant modes. The parameters are set
         as $a=1.6,b=2,c=1,\text{C}_1=3\times10^3,t=a^2/7$.}\label{tabResoVolMin}
\renewcommand\arraystretch{1.3}
\begin{tabular}
 {|l||c|c|c|c|}
  \hline
  $~$ & $m^2$ & $m$ & $\Gamma$ & $\tau$  \\
  \hline\hline
  $n=1$ &  $1.6762$  &  $1.2947$  &  $4.045\times10^{-10}$  &   $2.472\times10^9$    \\
  \hline
  $n=2$ &  $2.8490$  &  $1.6879$  &  $1.2745\times10^{-5}$  &   $7.846\times10^4$  \\
  \hline
  $n=3$ &  $3.5217$  &  $1.8766$  &  $0.0194$               &   $51.5282$         \\
  \hline
\end{tabular}
\end{center}
\end{table}

Then, in the case of parameter $t=a^2/6$, the effective potential belongs to the P\"{o}schl-Teller-like potential. The potential
well located at the brane position could localize a finite number of massive scalar modes on the thick brane. From
Eq. (\ref{LmtPTMinSca}), we can see that the limit $\text C$ is dependent on parameters. We will focus on the influence of
$\text C_1$ here. The shapes of the effective potential are shown in Fig. \ref{PTMin}, with $\text C_1=1\times10^3,2\times10^3$, and
$3\times10^3$. It can be seen that the potential well within the effective potential becomes deeper when $\text C_1$ is larger.
The localized massive modes can be solved through numerical methods, and the mass spectra are listed as follows:
\begin{eqnarray}
  m_n^2=&\{0,1.95,3.47,4.58,5.34,5.80\},\ \ \ \ \ \ \ \ \ \ \ \ \ \ \
  \ \ \ \ \ \ \ \ \ \ &\text{for}~\text C_1=1\times10^3,    \label{SpecPTMin1} \\
  m_n^2=&\{0,2.56,4.70,6.43,7.80,8.84,9.61,10.13,10.48\},\ \ &\text{for}~\text C_1=2\times10^3,    \label{SpecPTMin2} \\
  m_n^2=&\{0,3.01,5.60,7.79,9.60,11.08,12.26,13.17,~\ ~\ \ \ \ \       &\nonumber         \\
        &13.85,14.34,14.68\},\ \ \ \ \ \ \ \ \ \ \ \ \ \ \ \ \ \ \
       \ \ \ \ \ \ \ \ \ \ \ \ \ \ \ \ &\text{for}~\text C_1=3\times10^3.     \label{SpecPTMin3}
\end{eqnarray}
The number of the localized massive modes increases when $\text C_1$ is larger. Therefore, in the case of $t=a^2/6$, the scalar
zero mode, as well as a finite number of massive modes can be localized on the thick brane.
\begin{figure} [htbp]
\begin{center}
\includegraphics[width= 0.45\textwidth]{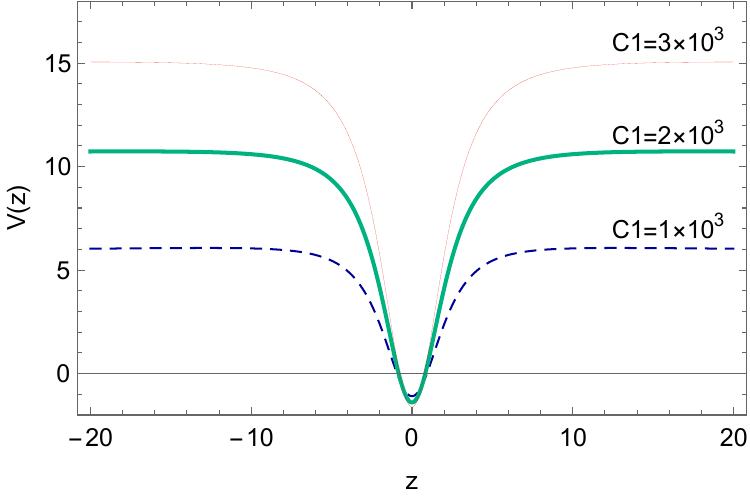}
\caption{The effective potential $V(z)$ with parameters
         $a=1.6,b=2,c=1,t=a^2/6$, and $\text{C}_1=1\times10^3,2\times10^3,3\times10^3$.}
\label{PTMin}
\end{center}
\end{figure}

Lastly, in the case of parameter $t>a^2/6$, there is a infinitely deep well, which will confines all massive KK modes
on the thick brane. For instance, we illustrate the effective potential and the mass spectra of lower localized massive
modes in Fig. \ref{SpecIDWMin} with parameters $a=1,b=2,c=1,\text{C}_1=2$, and $t=1$. In this case, all the KK modes, including
the scalar zero mode, are localized on the brane, forming a infinite discrete spectra of mass.
\begin{figure} [htbp]
\begin{center}
\includegraphics[width= 0.45\textwidth]{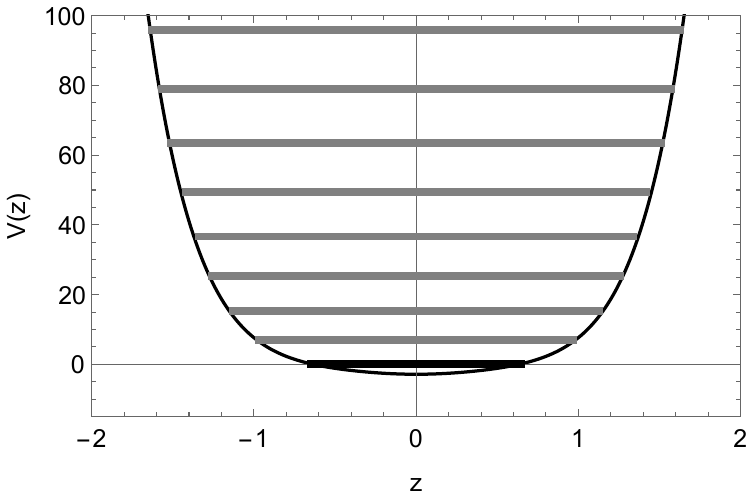}
\caption{The effective potential $V(z)$ with parameters $a=1,b=2,c=1,\text{C}_1=2,t=1$; the thick black line
         corresponds to $m_0^2=0$ and the first eight massive levels of the mass spectra are given
         by the grey lines.}
\label{SpecIDWMin}
\end{center}
\end{figure}    }

\vspace{0.5cm}
On the other hand, for the scenario of parameter $t<0$, by setting $V(z(y=0))=0$, we can obtain the following solutions
from Eq. (\ref{MinVzyt}):
\begin{eqnarray}
case~\text{I}: & &t=-\frac{a^2(10a^2+9)}{3(5a^2+4)}, \ \ \ \ \ \mathrm{C}_{1}=\frac{10a^2+9}{10a^2};\label{MinVzyt01}\\
 case~\text{II}: & &t=\frac{\mathrm{ProductLog}\bigg[\mathrm{C}_5,-\frac{a^2(10a^2+9)}{3(5a^2+4)}\ln(\frac{10a^2\mathrm{C}_{1}}{10a^2+9})\bigg]}
{\ln(\frac{10a^2\mathrm{C}_{1}}{10a^2+9})}, \nonumber\\
& &\mathrm{C}_{1}\neq\frac{10a^2+9}{10a^2}\text{ and } \mathrm{C}_5\in Integers,\label{MinVzyt02}
\end{eqnarray}
where ProductLog is the Lambert $W-$function.

For $case~\text{I}$, by setting $\mathrm{C}_{1}=\frac{9+10a^2}{10a^2}$, the potential
$V(z(y=0))$ becomes
\begin{eqnarray}\label{MinVzyt1}
 V(z(y=0))= -\frac32b^2e^{\frac{1}{3}a^2(1-6c)}\bigg(\frac{3(5a^2+4)}{10a^2+9}t+a^2\bigg),
\end{eqnarray}
which is a linear expression with respect to $t$. The effective potential for various values of $t$ is depicted in Fig. \ref{MinVt1}.
From this figure, we can see that the effective potential at $y=0$ decreases monotonously as $t$ increases.

For $case~\text{II}$, we present the shapes of $V(z(y=0))$ concerning the parameter $t$ in Fig. \ref{MinVt2} with parameters
$a=1,b=2,c=1$, and $\mathrm{C}_{1}=2$. Here, we only consider real values of $t$ for Eq. (\ref{MinVzyt02}), which
are $-95.79420$ and $-0.73057$. These two values correspond to $\mathrm{C}_5=-1$ and $0$, respectively.

Given the scalar zero mode is normalizable, a positive potential $V(z(y))$ at $y=0$ implies the existence of several 
negative wells, as the brane model is of $\mathbb{Z}_2$ symmetry. For example, Fig. \ref{figMinChi-V-t} displays the 
scalar zero mode and effective potential with parameters $\mathrm{C}_{1}=2, a=1, b=2,c=1$, and $t=-2$. Additionally, 
profiles for the case $t=0.1$ ($=\frac{a^2}{10}$) are provided for reference. The figures demonstrate that for $t=-2$, 
the scalar zero mode possesses a local minimum at the origin of the extra dimension, which differs from the global 
maximum for the case of $t=0.1$. From a quantum mechanics perspective, this local minimum implies that 4D scalar 
particles tend to approach the boundaries of the brane, with their probability densities being suppressed at $y=0$. 
Moreover, as previously mentioned, the effective potential $V(z(y))$ for $t=-2$ exhibits a double well structure. 
Two negative wells are symmetrically positioned at both sides of a barrier, which maintains a global maximum at the 
origin of the extra dimension.

\begin{figure}[htb]
\begin{center}
\subfigure[$case$ I]{\label{MinVt1}
\includegraphics[width= 0.45\textwidth]{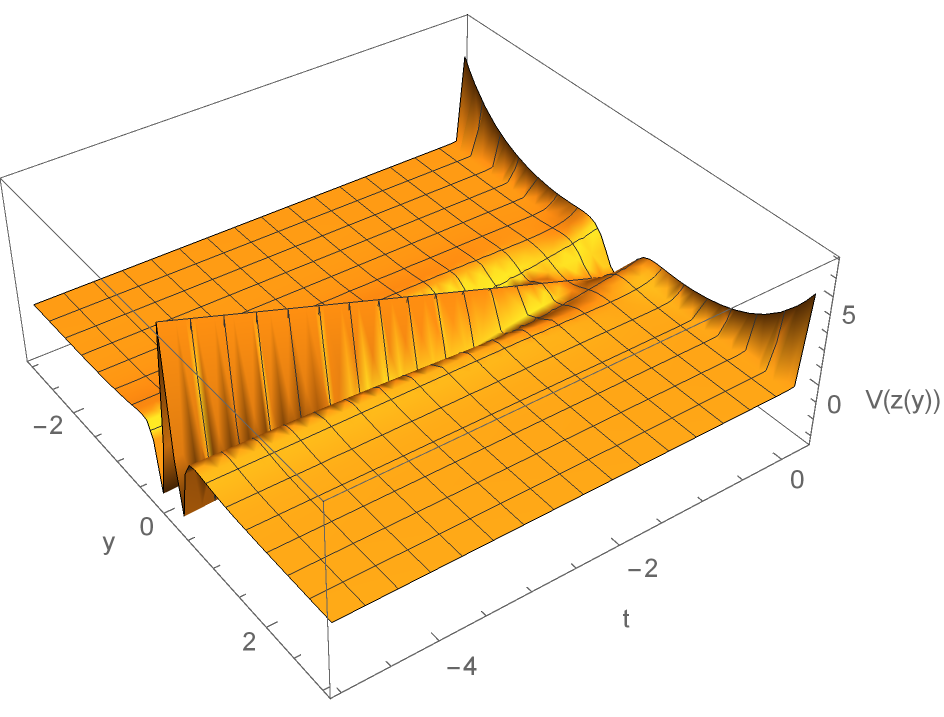}}
\subfigure[$case$ II]{\label{MinVt2}
\includegraphics[width= 0.45\textwidth]{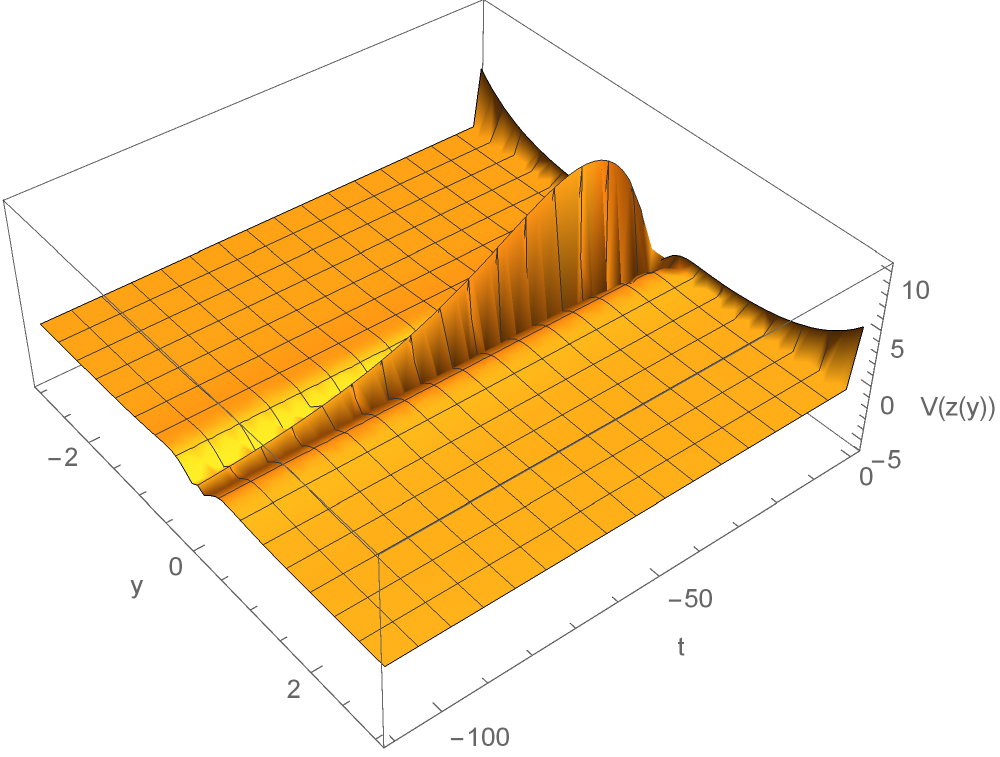}}
\end{center}\vskip -5mm
\caption{The shapes of the potential $V(z(y))$ for (a) $case$ I and (b) $case$ II.
 The parameters are set as $a=1$, $b=2$, $c=1$. And in Fig. (a), we set $\mathrm{C}_{1}=\frac{9+10a^2}{10a^2}$.
 In Fig. (b), we set $\mathrm{C}_{1}=2$. }
 \label{fig_V_Phi_phiz}
\end{figure}

\begin{figure} [htbp]
\begin{center}
\includegraphics[width= 0.45\textwidth]{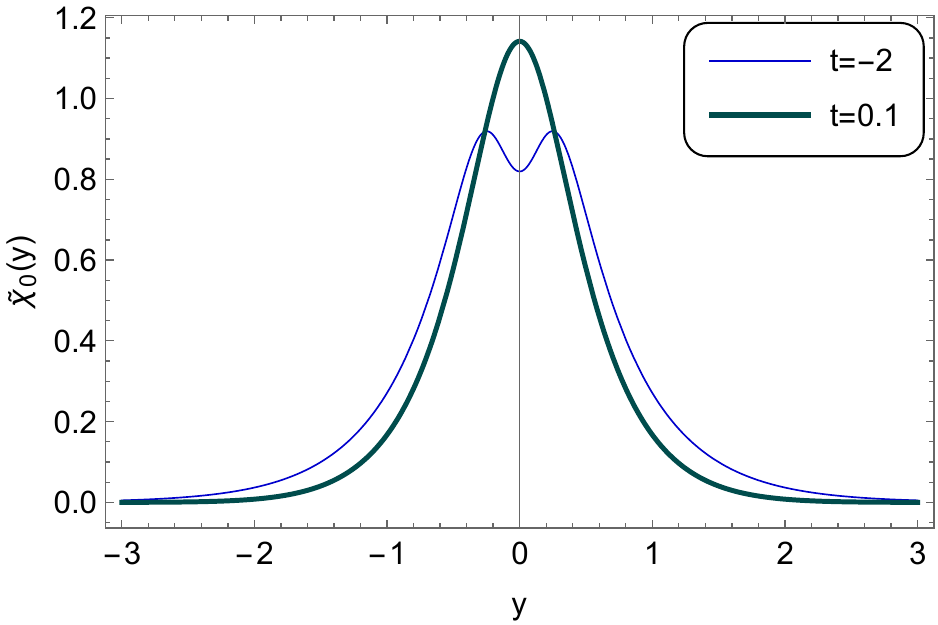}
\includegraphics[width= 0.45\textwidth]{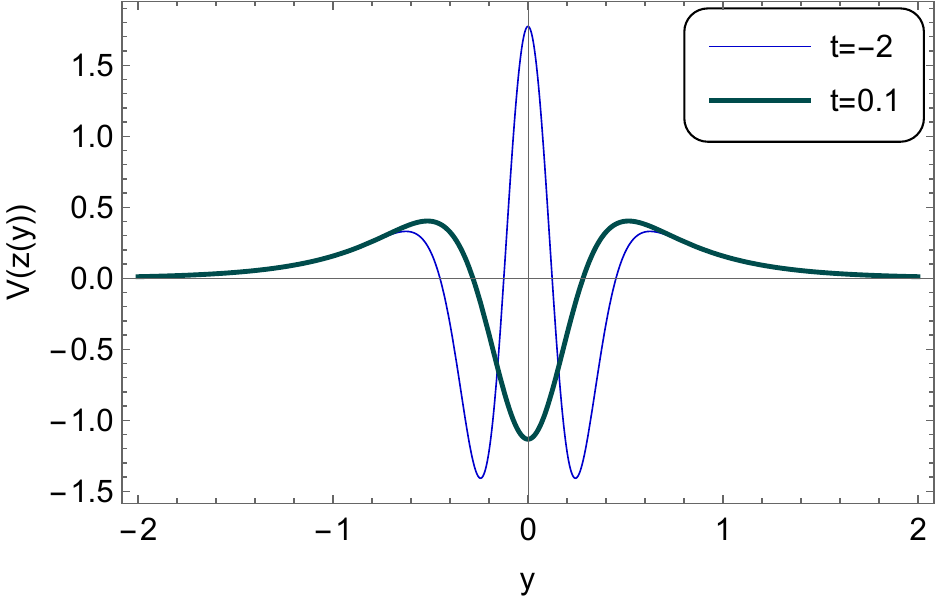}
\caption{The zero mode $\tilde{\chi}_0(y)$ and the effective potential
  $V(z(y))$ with parameters $a=1,b=2,c=1$ and $\mathrm{C}_{1}=2$. The
  coupling parameter is set as $t=-2$ for the thin line, and $t=0.1$
  for the thick line, respectively.}
\label{figMinChi-V-t}
\end{center}
\end{figure}

%%%%%%%%%%%%%%%%%%%%%%%%%%%%%%

\subsection{dS$_4$ Brane}\label{sec6}

{
For the dS brane case, the $3-$brane is warped and displays a positive scalar curvature. The warp factor of the brane
model is assumed as \cite{Wang_PRD}:
\begin{eqnarray}
 A(z)&=&-\delta\ln(\cosh(H\frac{z}{\delta})),    \label{dSbrane}   \\
 \varphi(z)&=&\sqrt{3\delta(1-\delta)}\arcsin(\tanh{\frac{Hz}{\delta}}),
\end{eqnarray}
where $\delta$ is a constant satisfying $0<\delta\leq2/3$, $H$ represents the dS parameter, and the 4D cosmological
constant is denoted as $\Lambda_4=3H^2$.

For this model, the effective potential $V_{\text{eff}}(z)$ (\ref{Veff}) for the perturbations of scalar can be expressed as
\begin{eqnarray}\label{veffb}
  V_{\text{eff}}(z) =-\frac{H^2(3\delta-2)[2+5\delta+3\delta(\text{sech}(H\frac{z}{\delta}))^2]}{4\delta^2}.
\end{eqnarray}
Based on this expression (\ref{veffb}), it is evident that when $0<\delta\leq2/3$, $V_{\text{eff}}(z)$ remains non-negative.
Consequently, there are no KK modes with negative $m^2$, and the thick brane model is stable under the scalar perturbations.
For example, the effective potential $V_{\text{eff}}(z)$ (\ref{veffb}) is plotted in Fig. \ref{figVeffScaPdS} with parameters
$\delta=0.5$, and $H=1$.
\begin{figure} [htbp]
\begin{center}
\includegraphics[width= 0.45\textwidth]{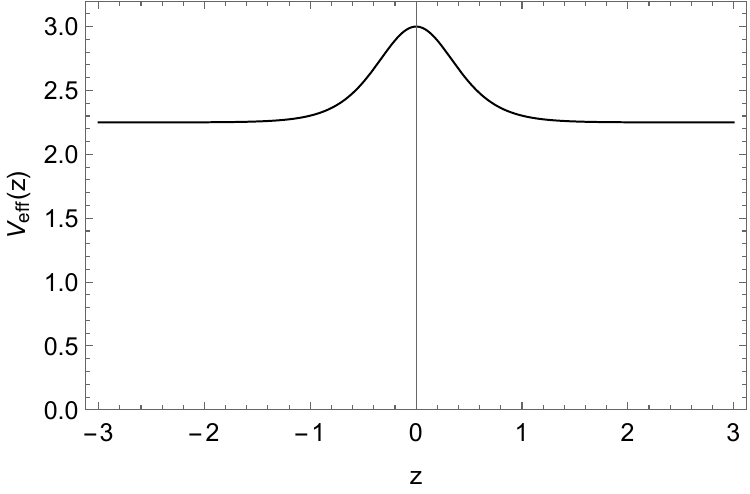}
\caption{{The effective potential $V_{\text{eff}}(z)$ (\ref{veffb}) corresponding to the stability
         under the scalar perturbations in dS$_4$ brane case. The parameters are set as
         $\delta=0.5$ and $H=1$.}}
\label{figVeffScaPdS}
\end{center}
\end{figure}

\vspace{0.1cm}
Since this brane model holds the $\mathbb{Z}_2$ symmetry, we will concentrate on the asymptotic behaviors
as $z\rightarrow+\infty$ for the subsequently quantities.

Based on the warp factor (\ref{dSbrane}), the scalar curvature (\ref{Rz}) becomes
\begin{eqnarray} \label{ScaCurvdS}
  R(z)=\frac{4H^2(3\delta+2)}{\delta}(\cosh(H\frac{z}{\delta}))^{2\delta-2}.
\end{eqnarray}
{From this expression, it can be seen that the 5D spacetime is asymptotically flat. Hence,
the coupling function $F(R)$ tends to value $1$ when far away from the brane.} In accordance with
Eq. (\ref{Vz}), the effective potential has the following asymptotic behaviors as $z\rightarrow+\infty$:
\begin{eqnarray} \label{VzminimdS}
  V(z\rightarrow+\infty)&\rightarrow&\frac{3}{2} A''(z)+\frac{9}{4}A'^{2}(z)           \nonumber     \\
    &=& \frac{3H^2}{4\delta}\bigg(3\delta-(3\delta+2)(\text{sech}(\frac{Hz}{\delta}))^2\bigg),
\end{eqnarray}
which is just the effective potential observed in the minimal coupling case. For this expression, we can further
find that the effective potential $V(z)$ tends to limit $9H^2/4$ at infinity.

Regarding the coupling function $F(R)$, we propose two specific forms, one of which is a polynomial and the other 
is an exponential function.
\vspace{0.2cm}

$case$ I: $F(R)=1+t_1R+t_2R^2$

For the given coupling function, both coupling parameters $t_1$ and $t_2$ are positive. In terms of Eq. (\ref{zeroz1}),
the expression for scalar zero mode $\tilde{\chi}_0(z)$ is
\begin{eqnarray} \label{1ZMdS}
  \tilde{\chi}_0(z) &&= N_2(\cosh(H\frac{z}{\delta}))^{-\frac32\delta}   \nonumber  \\
        &&\times\sqrt{1+\frac{4H^2(3\delta+2)(\cosh(H\frac{z}{\delta}))^{2\delta-2}(t_1\delta
     +4H^2t_2(3\delta+2)(\cosh(H\frac{z}{\delta}))^{2\delta-2})}{\delta^2}}.\;\;\;\;
\end{eqnarray}
Since the parameter $0<\delta\leq2/3$, the factors involving the hyperbolic cosine function in this expression always
approach zero as $z\rightarrow+\infty$. Consequently, the scalar zero mode satisfies the localization condition
(\ref{LocKK}), and can be localized on the thick brane.

For the effective potential (\ref{VzminimdS}), it always tends to $9H^2/4$ when $z\rightarrow+\infty$. However,
it exhibits different behavior at finite positions along the extra dimension. For instance, Fig. \ref{fig1FREffPotdS}
depicts the effective potential $V(z)$ with specific parameter values. As illustrated in the figures, the
asymptotic behaviors of the effective potential remain unchanged, reaching $9H^2/4$ as $z\rightarrow+\infty$.
Moreover, symmetrically positioned potential barriers emerge on both sides of the brane. These barriers
gradually widen with increasing $t_2$, and exhibit a slight increase with lower values of $t_1$ and higher
values of $t_2$.
\begin{figure}[htb]
\begin{center}
\subfigure[$t_2=5$]{\label{11FREffPotdS}
\includegraphics[width= 0.45\textwidth]{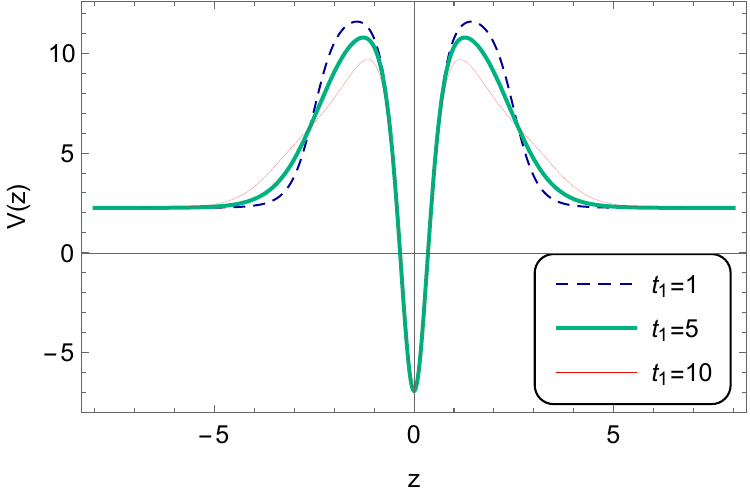}}
\subfigure[$t_1=1$]{\label{12FREffPotdS}
\includegraphics[width= 0.45\textwidth]{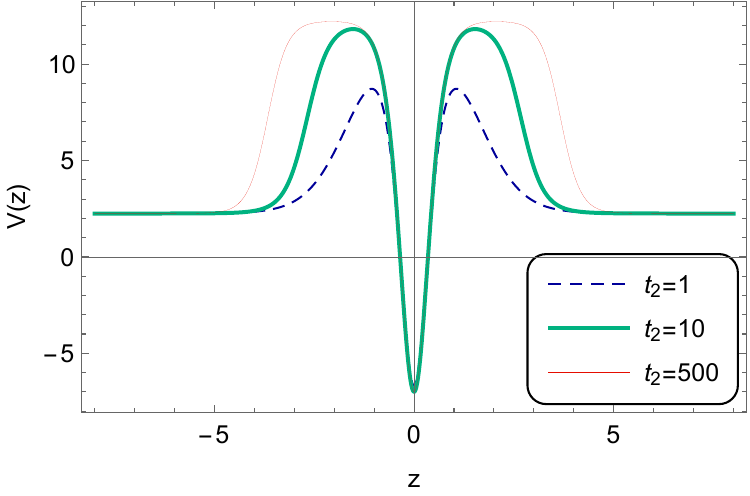}}
\end{center}\vskip -5mm
\caption{The effective potential
      $V(z)$ with parameters $\delta=0.5,H=1,t_2=5$ and $t_1=1,5,10$ in (a), and with parameters
      $\delta=0.5,H=1,t_1=1$ and $t_2=1,10,500$ in (b).}
 \label{fig1FREffPotdS}
\end{figure}

The potential barriers within the effective potential shown in Fig. \ref{fig1FREffPotdS} are higher than the
potential limit. The resonant KK modes could exist in this case. We depict the profiles of the relative probability
for the scalar resonant mode in Fig. \ref{1ResoSpecPFRdSSca}, utilizing
parameters $\delta=0.5,H=1,t_1=1$, and $t_2=10$. The corresponding effective potential is also displayed in the same figure.
In the mass spectra, the localized zero mode is the ground state, and the massive mode corresponds to the resonant mode.
Thus, by introducing the coupling factor $F(R)$, it is feasible to localize the zero mode, and quasi-localize the massive 
modes on the thick brane.
\begin{figure}[htb]
\begin{center}
\subfigure[$V(z),m^2$]{\label{1ResoSpecFRdSSca}
\includegraphics[width= 0.45\textwidth]{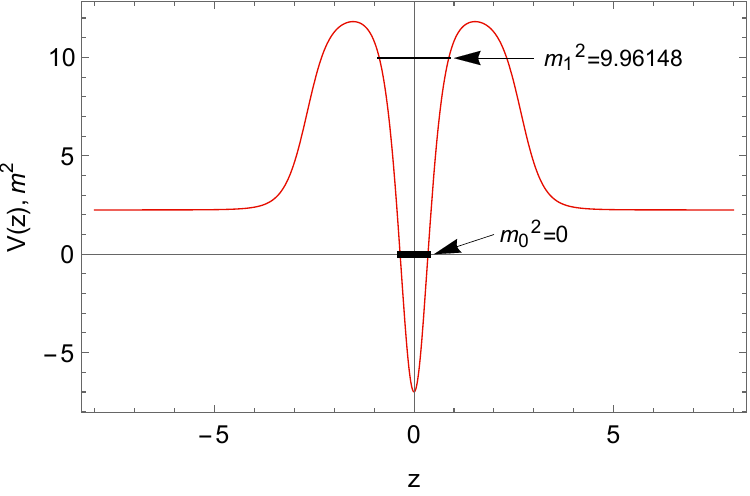}}
\subfigure[$P$]{\label{1ResonancePdS}
\includegraphics[width= 0.45\textwidth]{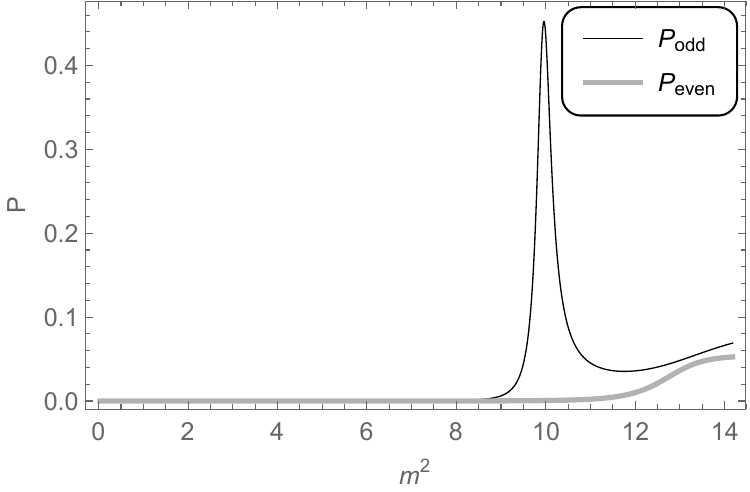}}
\end{center}\vskip -5mm
\caption{(a) The effective potential $V(z)$ and the mass spectra.
         (b) The corresponding relative probability $P$ of the resonant modes. The parameters
         are set as $\delta=0.5,H=1,t_1=1$, and $t_2=10$. }
 \label{1ResoSpecPFRdSSca}
\end{figure}

\vspace{0.5cm}
$case$ II: $F(R)=e^{t_1(1-e^{t_2R})}$

Concerning this coupling function, both parameters $t_1$ and $t_2$ are positive. Then, the scalar
zero mode $\tilde{\chi}_0(z)$ reads
\begin{eqnarray} \label{3ZMdS}
  \tilde{\chi}_0(z) = N_2(\cosh(H\frac{z}{\delta}))^{-\frac32\delta}
      \sqrt{e^{t_1-t_1e^{\frac{4H^2t_2(3\delta+2)}{\delta}(\cosh(H\frac{z}{\delta}))^{2\delta-2}}}}.
\end{eqnarray}

Furthermore, the localization condition for scalar zero mode is
\begin{eqnarray} \label{3LocCondZMdS}
  \int \tilde{\chi}_0^2(z)dz = N_2^2\int (\cosh(H\frac{z}{\delta}))^{-3\delta}
      e^{t_1-t_1e^{\frac{4H^2t_2(3\delta+2)}{\delta}(\cosh(H\frac{z}{\delta}))^{2\delta-2}}}dz<\infty.
\end{eqnarray}
{It can be seen that with $0<\delta\leq2/3$, this condition is met, and the scalar zero mode is
normalizable.

Additionally, at position $z=0$, the scalar zero mode (\ref{3ZMdS}) and its lower-order derivatives become
\begin{eqnarray}
  \tilde{\chi}_0(z=0) &=& N_2\sqrt{e^{t_1-t_1e^{\frac{4H^{2}(3\delta+2)t_{2}}{\delta}}}},       \label{3ZMOridS}      \\
  \partial_z\tilde{\chi}_0(z=0) &=& 0,                                   \label{31stDeriZMOridS}    \\
  \partial^2_z\tilde{\chi}_0(z=0) &=& -\frac{H^2}{2\delta^3}\sqrt{e^{t_1-t_1e^{\frac{4(3\delta+2)H^2t_2}{\delta}}}}    \nonumber   \\
    & & \times\bigg(3\delta^2+8H^2(\delta-1)(3\delta+2)t_1t_2e^{\frac{4H^2(3\delta+2)t_2}{\delta}}\bigg).       \label{32ndDeriZMOridS}
\end{eqnarray}
From Eq. (\ref{3ZMOridS}), it can be seen that the scalar zero mode tends to zero at $z=0$ as parameter $t_2$
increases. As the zero mode (\ref{3ZMdS}) is normalizable, we can infer that it is localized on both sides of 
the origin of extra dimension.

For Eq. (\ref{31stDeriZMOridS}), it implies that the zero mode could take a local extremum at position $z=0$. Concerning
the type of this extremum, we will further analyze the sign of the second-order derivative (\ref{32ndDeriZMOridS}).
By setting $\partial^2_z\tilde{\chi}_0(z=0)=0$, we can get the critical values for parameters $t_1$ and $t_2$:
\begin{eqnarray}
  t_{1\text C} &=& -\frac{3\delta^2}{8H^2t_2(3\delta+2)(\delta-1)}e^{-\frac{4H^2(3\delta+2)t_2}{\delta}},      \label{t1Ze32ndDeriZMOridS}    \\
  t_{2\text C} &=& \frac{\delta}{4H^2(3\delta+2)}\text{ProductLog}[\text{C}_2,-\frac{3\delta}{2t_1(\delta-1)}],   \label{t2Ze32ndDeriZMOridS}
\end{eqnarray}
where $\text{C}_2$ is an integer. Referring to Eq. (\ref{32ndDeriZMOridS}), it can be concluded that if parameters
$t_1$ or $t_2$ fall below their respective critical values, the second-order derivative (\ref{32ndDeriZMOridS}) is
negative, and the zero mode reaches a local maximum at position $z=0$. When parameters $t_1$ or $t_2$ are set as their
critical values, the second-order derivative (\ref{32ndDeriZMOridS}) equals zero, and there will be a plateau,
instead of a local extremum for zero mode at position $z=0$. Finally, if parameters $t_1$ or $t_2$ are larger than
their respective critical values, the second-order derivative (\ref{32ndDeriZMOridS}) becomes positive, and the zero mode
takes a local minimum at position $z=0$.

The plots of the scalar zero mode $\tilde{\chi}_0(z)$ are presented in Fig. \ref{++ZMEffPFRdSSca} with specific
values of parameters. As these figures illustrate, the single peak shaping the zero mode splits into two when
parameters $t_1$ or $t_2$ exceed their respective critical values, indicating the emergence of a split zero mode.
Subsequently, as parameters $t_1$ and $t_2$ continue to increase, these two peaks gradually move away from the
origin, and the splitting becomes more pronounced.

\begin{figure}[htb]
\begin{center}
\subfigure[$\tilde{\chi}_0(z)$ for $t_1=1$]{\label{t2++3ZMFRdSSca}
\includegraphics[width= 0.42\textwidth]{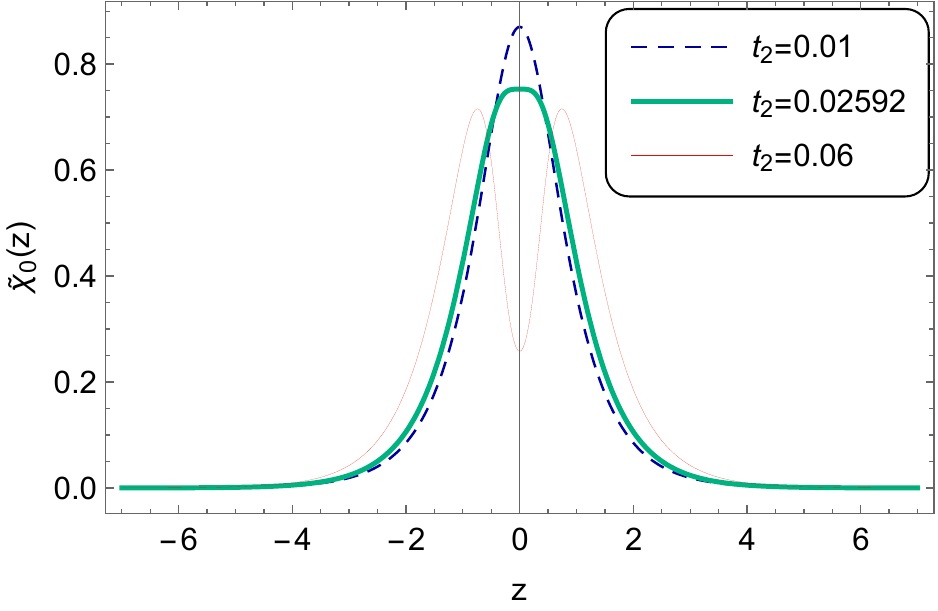}}
\subfigure[$V(z)$ for $t_1=1$]{\label{t2++3EffPotFRdSSca}
\includegraphics[width= 0.42\textwidth]{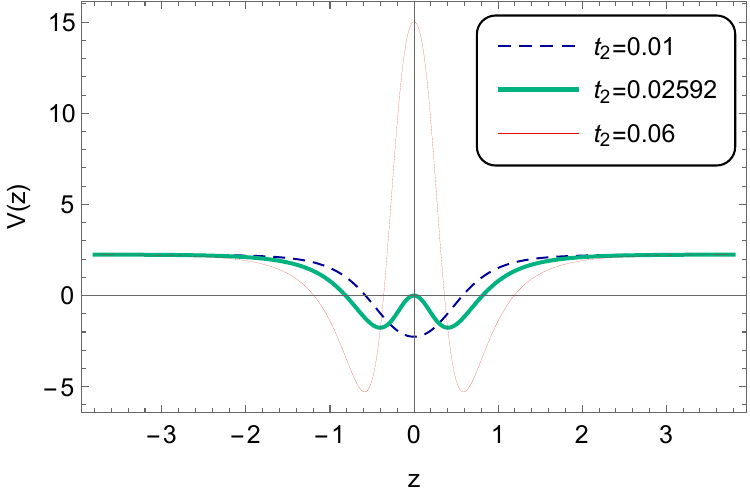}}
\subfigure[$\tilde{\chi}_0(z)$ for $t_2=0.06$]{\label{t1++3ZMFRdSSca}
\includegraphics[width= 0.42\textwidth]{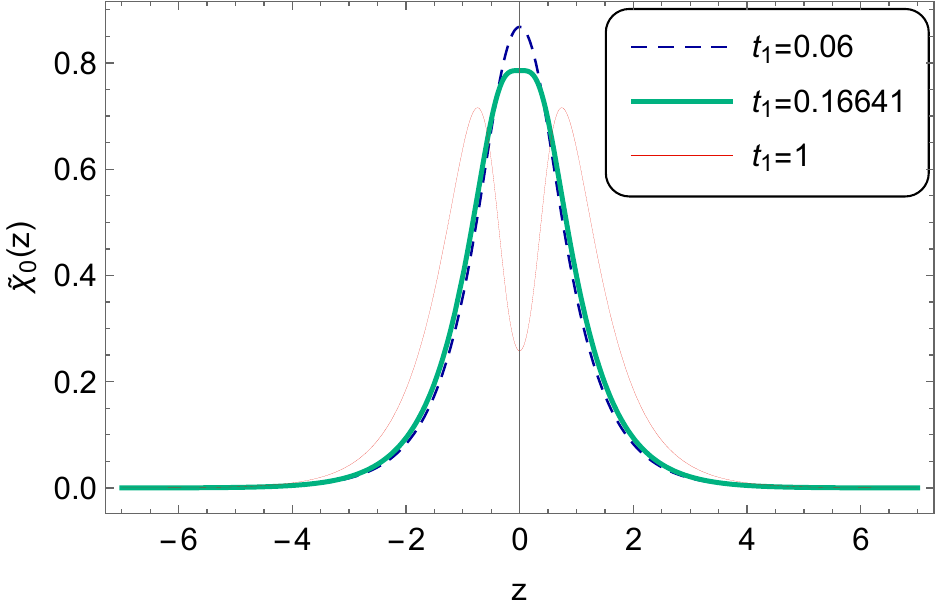}}
\subfigure[$V(z)$ for $t_2=0.06$]{\label{t1++3EffPotFRdSSca}
\includegraphics[width= 0.42\textwidth]{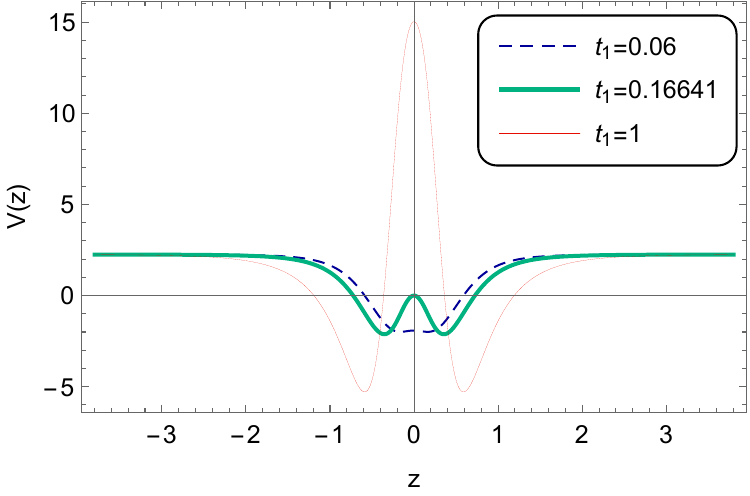}}
\end{center}\vskip -5mm
\caption{The scalar zero mode and the effective potential for KK modes. The parameters
         are set as $\delta=0.5,H=1$. %$t_1<t_{1\text C}$ ((c) and (d)) and $t_2<t_{2\text C}$ ((a) and (b)) for the dashed line.
         %$t_1=t_{1\text C}$ ((c) and (d)) and $t_2=t_{2\text C}$ ((a) and (b)) for the thick line. $t_1>t_{1\text C}$ ((c) and (d)) and
         %$t_2>t_{2\text C}$ ((a) and (b)) for the thin line.}
         {The dashed line represents parameters $t_1<t_{1\text C}$ or $t_2<t_{2\text C}$.
         The thick line represents parameters $t_1=t_{1\text C}$ or $t_2=t_{2\text C}$.
         The thin line represents parameters $t_1>t_{1\text C}$ or $t_2>t_{2\text C}$.}}
 \label{++ZMEffPFRdSSca}
\end{figure}

For the effective potential of scalar KK modes, Eq. (\ref{VzminimdS}) describes its asymptotic behavior 
$V(z\rightarrow+\infty)\rightarrow9H^2/4$, indicating that the coupling returns to the minimal coupling 
at infinity. Moreover, at position $z=0$, the effective potential becomes
\begin{eqnarray} \label{3EffPotOridS}
  V(z=0)=-\frac{H^2}{2\delta^3}\bigg(3\delta^2 +8H^2(\delta-1)(3\delta+2)t_1t_2e^{\frac{4H^2(3\delta+2)t_2}{\delta}}\bigg).
\end{eqnarray}
From this expression, we observe that for $0<\delta\leq2/3$, the effective potential equals $-3H^2/(2\delta)$ when 
either $t_1$ or $t_2$ are zero, and it increases with the values of $t_1$ and $t_2$. Furthermore, this expression 
shares a common factor with Eq. (\ref{32ndDeriZMOridS}). Consequently, if either $t_1$ or $t_2$ are set to their 
respective critical values, the effective potential will reach zero at $z=0$, as well as the second-order derivative 
of the scalar zero mode. This result aligns with the Schr\"{o}dinger-like equation (\ref{eq3}), which is evident  
when the latter is rewritten as
\begin{eqnarray}  \label{SchroEqdS}
  -\partial^2_z\tilde{\chi}_n(z)=(m^2_n-V(z))\tilde{\chi}_n(z).
\end{eqnarray}
For the zero mode $m_0=0$, this expression becomes: 
\begin{eqnarray}  \label{ZMSchroEqdS}
  \partial^2_z\tilde{\chi}_0(z)=V(z)\tilde{\chi}_0(z).
\end{eqnarray}
At position $z=0$, if $t_1=t_{1\text C}$ or $t_2=t_{2\text C}$, the expression within the l.h.s of the 
above equation equals zero, resulting in $V(z)=0$.

The profiles of the effective potential for scalar KK modes are also depicted in  Fig. \ref{++ZMEffPFRdSSca} with the
same parameter values. As shown in these figures, a barrier emerges at $z=0$ when the parameters $t_1$ or $t_2$ are
greater than their respective critical values, and the height of this barrier increases as parameters $t_1$ and $t_2$
rise. Additionally, two potential wells symmetrically appear adjacent to this barrier, localizing the scalar zero
mode on both sides of the origin.

Further insights into the effective potential of KK modes can be obtained by increasing the parameter $t_2$.
With a larger $t_2$, the single local maximum of the barrier splits into two, creating a potential well
with a positive lower boundary at $z=0$. This configuration allows for the existence of resonant KK modes.
For example, the mass spectra and relative probability for resonant modes are visually depicted in
Fig. \ref{++SpecP3ResoEffPotFRdSScA} with parameters $\delta=0.5,H=1,t_1=1$, and $t_2=0.14$. In
Fig. \ref{++Spec3ResoEffPotFRdSScA}, the localized zero mode represents the ground state, and four massive
modes are resonant modes. Detailed information regarding the mass, width, and lifetime of all scalar resonant
KK modes is provided in Table \ref{tabdS}.
\begin{figure}[htb]
\begin{center}
\subfigure[$V(z),m^2$]{\label{++Spec3ResoEffPotFRdSScA}
\includegraphics[width= 0.46\textwidth]{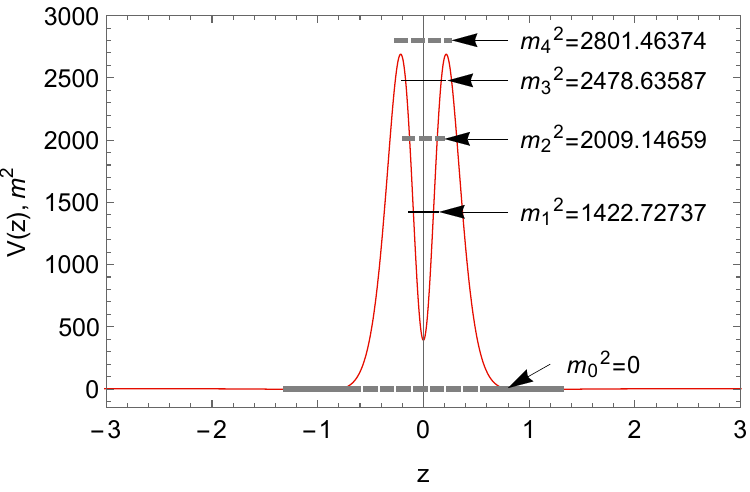}}
\subfigure[$P$]{\label{++P3ResoEffPotFRdSScA}
\includegraphics[width= 0.45\textwidth]{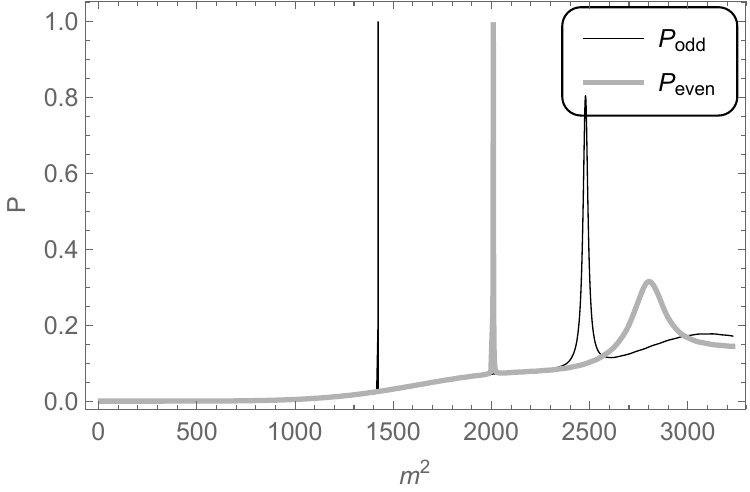}}
\end{center}\vskip -5mm
\caption{(a) The effective potential $V(z)$ and the mass spectra.
         (b) The corresponding relative probability $P$. The parameters
         are set as $\delta=0.5,H=1,t_1=1$, and $t_{2}=0.14$.}
 \label{++SpecP3ResoEffPotFRdSScA}
\end{figure}

\begin{table}[ht]
\begin{center}
\caption{The mass, width, and lifetime for
         scalar resonant modes. The parameters are set as $\delta=0.5,
         H=1,t_1=1$ and $t_{2}=0.14$.}\label{tabdS}
\renewcommand\arraystretch{1.3}
\begin{tabular}
 {|l||c|c|c|c|}
  \hline
  $~$ & $m^2$ & $m$ & $\Gamma$ & $\tau$  \\
  \hline\hline
  $n=1$ &  $1422.727$  &  $37.7191$  &  $8.175\times10^{-4}$  &   $1.223\times10^3$   \\
  \hline
  $n=2$ &  $2009.147$  &  $44.8235$  &  $0.0268$              &   $37.2518$  \\
  \hline
  $n=3$ &  $2478.636$  &  $49.7859$  &  $0.3376$              &   $2.9625$    \\
  \hline
  $n=4$ &  $2801.464$  &  $52.9289$  &  $3.6986$              &   $0.2704$    \\
  \hline
\end{tabular}
\end{center}
\end{table}

Figure \ref{++Spec3ResoEffPotFRdSScA} showcases that the scalar zero mode is localized on both sides of the origin
of the extra dimension, while the massive KK modes are quasi-localized at the origin. This distribution of scalar
KK modes describes a separation of scalars along the extra dimension. The shapes of probability densities for the
zero mode and the four resonant KK modes are depicted in Fig. \ref{figProbDenChidS}. The probability density of scalar
zero mode forms two lumps at both sides of the origin, and the probability densities for the resonant KK modes
center at the origin. In Refs. \cite{Arkani-HamedPRD2000,Arkani-HamedPRD2000-2,DCCaiPRD2006}, the so-called
"split fermion" model is proposed, where different fermion KK modes could be localized at different positions.
Here, the scalar zero mode and the massive modes can also be localized or quasi-localized at the different positions
of extra dimension.
\begin{figure}[htb]
\begin{center}
\includegraphics[width= 0.96\textwidth]{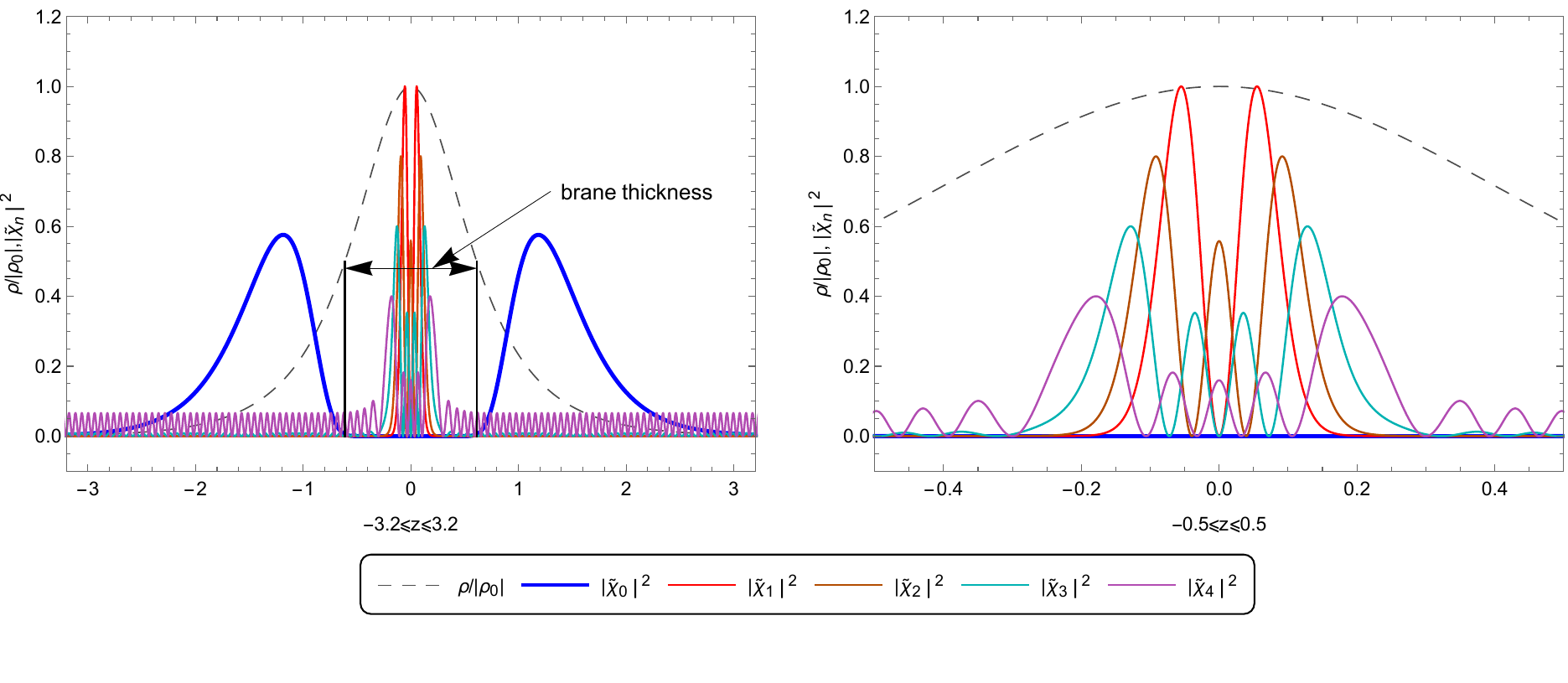}
\end{center}\vskip -2mm
\vspace{-0.2cm}
\caption{The probability densities for the scalar zero mode and the resonant KK modes. The parameters are set as $\delta=0.5,
         H=1,t_1=1$, and $t_{2}=0.14$.}
 \label{figProbDenChidS}
\end{figure}

Moreover, considering the energy density \cite{Wang_PRD}
\begin{eqnarray} \label{EnergDendS}
  \rho=\frac12((\bigtriangledown\varphi)^2+2V(\varphi)),
\end{eqnarray}
we plot the relative energy density $\rho/|\rho_0|$ with $\rho_0=\rho(z=0)$ in Fig. \ref{figProbDenChidS} using the same
parameter values. As usual, the thickness of the brane is defined as the full width at half maximum of the peak of the
energy density. From Fig. \ref{figProbDenChidS}, we observe that the two peaks of the probability density for the zero mode are
positioned on both sides of the brane. Besides, the four massive KK modes are quasi-localized on the brane. Thus, as the parameters
$t_1$ and $t_2$ increase, the height of the barrier at the origin rises, causing more massive KK modes to become quasi-localized on
the brane, and the two peaks of the probability density for the zero mode to gradually move further away from the origin.}

Therefore, by introducing the coupling factor $F(R)$, in addition to the localized scalar zero mode, quasi-localized
massive modes can also exist on the thick dS brane. Besides, with a specific form of coupling factor $F(R)$, the
zero mode can be localized on both sides of the brane.

%%%%%%%%%%%%%%%%%%%%%%%%%%%%%%

\subsection{AdS$_4$ Brane}\label{sec7}

In this section, we will discuss the localization of the scalar field in the case of an AdS$_4$ brane. The warp factor considered
here is of the form \cite{Wang_PRD,LiuPRD1101.4145}
\begin{eqnarray}
 A(z)  &=& -\delta \ln|\cos(\frac{H}{\delta}z)|,    \label{eAAdS}  \\
 \varphi(z) &=& \sqrt{3\delta(\delta-1)}\text{arcsinh}(\tan(\frac{H}{\delta}z)),   \label{eAphiAdS}
\end{eqnarray}
where parameter $\delta>1$. The domain of the extra dimension is $-z_b\leq z\leq z_b$ with $z_b=|\frac{\delta\pi}{2H}|$.
The metric with this warp factor (\ref{eAAdS}) has a naked singularity at $\pm z_b$. This singularity is similar to the one in Ref \cite{GremmPLBPRD}
and the one encountered in the AdS flow to $N=1$ super Yang-Mills theory \cite{GirardelloNP}. Gremm supported that this condition can
be resolved either by lifting the five-dimensional to ten dimensions or by string theory \cite{GremmPLBPRD}.

Concerning the stability of this brane model under the scalar perturbations, the potential $V_{\text{eff}}(z)$ (\ref{Veff}) becomes
\begin{eqnarray}
\label{VeffAdS}
 V_{\text{eff}}&=&
  -\frac{5}{2}A''+\frac{9}{4}A'^{2}+A'\frac{\varphi''}{\varphi'}
  -\frac{\varphi'''}{\varphi'}+2\left(\frac{\varphi''}{\varphi'} \right)^{2}
   +6H^2                    \nonumber      \\
   &=& \frac{H^2}{4\delta^2}(3\delta-2)(5\delta+2+3\delta(\sec(\frac{Hz}{\delta}))^2).
\end{eqnarray}
In this case, it is known that Eq. (\ref{4thEqScaPertMin_b}) can be solved with suitable harmonic functions.
There is the Breitenlohner-Freedman bound which allows the tachyonic mass to some extent from the condition of the
normalization \cite{PRD_koyama,BreitenlohnerPLB115,BreitenlohnerAP115}. The mass $m$ is bounded as
\begin{eqnarray} \label{MScaPertAdS}
  m^2\geq-\frac94,
\end{eqnarray}
which means that even when there are solutions with $-\frac94\leq m^2<0$, such solutions are stable in spite of the
tachyonic mass. Then, for the potential (\ref{VeffAdS}) in Eq. (\ref{4thEqScaPertMin_b}), further analysis can show that
the thick AdS brane is stable under the scalar perturbations \cite{LiuPRD1101.4145}.
\vspace{0.2cm}

Based on the warp factor (\ref{eAAdS}), the scalar curvature can be obtained:
\begin{eqnarray} \label{RAdS}
  R=-\frac{4H^2(3\delta+2)}{\delta}\cos(\frac{Hz}{\delta})^{2\delta-2}.
\end{eqnarray}
{It can be seen that the 5D scalar curvature remains negative, and converges to zero at the boundaries of the 
extra dimension, so the 5D spacetime is asymptotically flat. According to the second rule for the function $F(R)$, it  
tends to value $1$ at the boundaries of extra dimension.} In this AdS brane case, we consider the coupling factor of the 
form
\begin{eqnarray} \label{fRAdS}
  F(R)=e^{t_1(1-e^{-t_2R})},
\end{eqnarray}
where both parameters $t_1$ and $t_2$ are positive.

For a free scalar field with action (\ref{action_scalar}), the coupling reverts to the minimal coupling as
$z\rightarrow\pm\frac{\delta\pi}{2H}$. Introducing the coupling factor $F(R)$ does not alter the localization result of 
the scalar zero mode. In Ref. \cite{LiuPRD1101.4145}, it was demonstrated that for free scalar fields, the ground state is
a massive mode, and all the KK modes are bound states except for the zero mode, which cannot be localized on the brane.
Therefore, in the case under consideration, despite the existence of the coupling factor $F(R)$, the scalar zero mode
cannot be localized.

On the other hand, as analyzed in Sec. \ref{LocGF}, the inclusion of the coupling potential $V(\Phi, \varphi)$ could potentially
influence the localization of the scalar zero mode. Thus, both the coupling potential $V(\Phi,\varphi)$ and the coupling factor
$F(R)$ will be considered for this AdS brane case. The 5D action of a scalar field is assumed as
\begin{eqnarray}\label{actionAdS}
S=-\frac{1}{2}\int d^{5}x\sqrt{-g}~
          [F(R)g^{MN}\partial_{M}\Phi\partial_{N}\Phi-V(\Phi,\varphi)].
\end{eqnarray}

Furthermore, the Schr\"{o}dinger-like equation for scalar KK modes $\tilde\chi_n(z)$ can be obtained from action (\ref{actionAdS}):
\begin{eqnarray} \label{SchroAdS}
\left[-\partial^{2}_z+ V(z)\right]{\tilde{\chi}}_n(z)
  =m_{n}^{2} {\tilde{\chi}}_{n}(z),
\end{eqnarray}
where the effective potential $V(z)$ is
\begin{eqnarray} \label{VzAdS}
  V(z)=\frac{3}{2} A''(z)+\frac{9}{4}A'^{2}(z)
       +\frac{F''(R)}{2 F(R)}+\frac{3 A'(z) F'(R)}{2 F(R)}
       -\frac{F'^2(R)}{4 F^2(R)}+\frac{e^{2A}U(\varphi)}{F(R)}.
\end{eqnarray}
Function $U(\varphi)$ is defined by
\begin{eqnarray} \label{defUAdS}
  \frac{\partial V(\Phi,\varphi)}{\partial\Phi}=U(\varphi)\Phi+\mathcal{O}(\Phi^3),
\end{eqnarray}
and is calculated from Eq. (\ref{coupling potential}) as
\begin{eqnarray} \label{UAdS}
  U(\varphi)=2(\lambda\varphi^2-u^2).
\end{eqnarray}

Moving forward, we will discuss the localization of scalar KK modes in two cases: with or without the existence of the coupling
potential $V(\Phi,\varphi)$.

Firstly, if we omit the coupling potential $V(\Phi,\varphi)$ with $\lambda=0,u=0$, Eq. (\ref{SchroAdS}) reduces to Eq. (\ref{eq3}),
and the solution for the scalar zero mode (\ref{zeroz1}) remains valid, which becomes
\begin{eqnarray} \label{zero1AdS}
\tilde{\chi}_0(z)=e^{\frac12\bigg(1-e^{\frac{4H^2(3\delta+2)t_2(\cos(\frac{Hz}{\delta}))^{2\delta+2}}{\delta}}\bigg)t_1}
        (\cos(\frac{Hz}{\delta}))^{-\frac{3\delta}{2}}.
\end{eqnarray}
It is evident that this scalar zero mode diverges as $z\rightarrow\pm\frac{\delta\pi}{2H}$ and remains non-normalized.

Considering the massive KK modes, we will focus on Eq. (\ref{eq3}). Based on the brane model (\ref{eAAdS}), the coupling
turns back to minimal coupling as $z\rightarrow\pm\frac{\delta\pi}{2H}$, so the effective potential of KK modes exhibits 
the following asymptotic behaviors:
\begin{eqnarray} \label{AsyEff1AdS}
  V(z\rightarrow\pm\frac{\delta\pi}{2H})&\rightarrow & \frac32A''(z)+\frac{9}{4}A'^2(z)   \nonumber    \\
        &=& -\frac94H^2+\frac{3H^2(3\delta+2)}{4\delta}(\sec(\frac{Hz}{\delta}))^2.
\end{eqnarray}
It is evident that the effective potential diverges at the boundaries of the extra dimension. Consequently, all massive KK modes can be
localized on the brane. For instance, we depict the scalar zero mode and the effective potential in Fig. \ref{1IDWAdS} with
parameters $H=1$, $\delta=2$, $\lambda=0$, $u=0$, $t_1=1$, and $t_2=0.1$. The mass spectra of lower localized massive modes
are also displayed in the same figure. This result is consistent with the observations made in Ref. \cite{LiuPRD1101.4145}.
\begin{figure}[htb]
\begin{center}
\subfigure[$\tilde{\chi}_0(z)$]{\label{ZM1AdS}
\includegraphics[width= 0.45\textwidth]{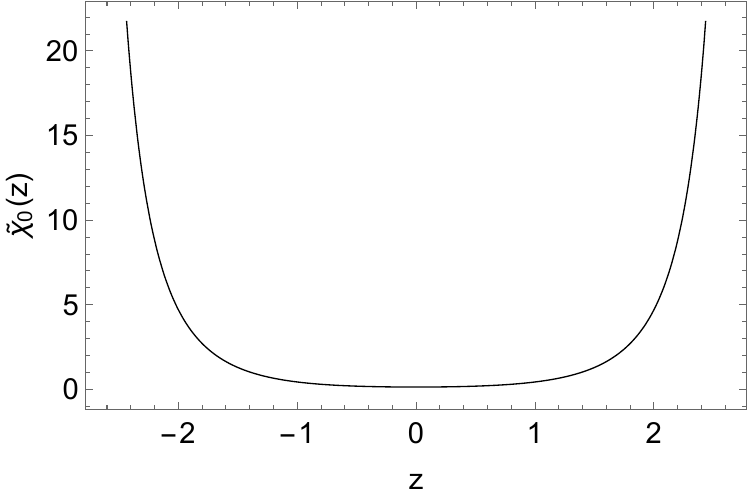}}
\subfigure[$V(z)$]{\label{Spec1IDWAdS}
\includegraphics[width= 0.45\textwidth]{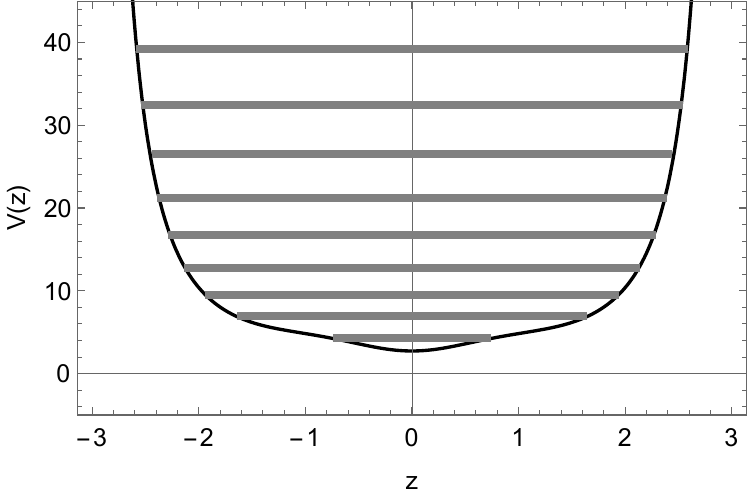}}
\end{center}\vskip -5mm
\caption{The scalar zero mode $\tilde{\chi}_0(z)$ and the effective potential $V(z)$ with parameters
         $H=1,\delta=2,\lambda=0,u=0,t_1=1,t_2=0.1$; the thick grey lines correspond to the first nine
         massive levels of the mass spectra.}
 \label{1IDWAdS}
\end{figure}

\vspace{0.2cm}
Then, we consider both the coupling potential $V(\Phi,\varphi)$ and the coupling factor $F(R)$ simultaneously. In this scenario,
factorizing the Schr\"{o}dinger-like equation (\ref{SchroAdS}) becomes challenging, and no analytical solution exists for the
scalar zero mode. Therefore, we will employ numerical methods to solve for both the zero mode and massive modes with Eq. (\ref{SchroAdS}).

As $z\rightarrow\pm\frac{\delta\pi}{2H}$, the factor $F(R)$ tends to value 1, and the effective potential (\ref{VzAdS}) has the
following asymptotic behaviors:
\begin{eqnarray} \label{AsyEff2AdS}
  V(z\rightarrow\pm\frac{\delta\pi}{2H})&\rightarrow & \frac32A''(z)+\frac{9}{4}A'^2(z)+e^{2A}U(\varphi)   \nonumber    \\
        &=& -2\bigg(u^2-3(\delta-1)\delta\lambda(\text{arcsinh}(\tan(\frac{Hz}{\delta})))^2\bigg)\cos(\frac{Hz}{\delta})^{-2\delta}    \nonumber    \\
        & & +\frac{3H^2}{2\delta}(\sec(\frac{Hz}{\delta}))^2+\frac94H^2(\tan(\frac{Hz}{\delta}))^2.
\end{eqnarray}
On the r.h.s. of this expression, the factor $\cos(\frac{Hz}{\delta})^{-2\delta}$ exhibits the highest order, and the factor
$-2(u^2-3(\delta-1)\delta\lambda(\text{arcsinh}(\tan(\frac{Hz}{\delta})))^2)$ determines the sign. Since the factor
$(\text{arcsinh}(\tan(\frac{Hz}{\delta})))^2$ diverges to positive infinity as $z\rightarrow\pm\frac{\delta\pi}{2H}$, the further
asymptotic behaviors of the effective potential depend on the parameter $\lambda$. If $\lambda>0$, the effective potential tends to
positive infinity at the boundaries of the extra dimension, and if $\lambda<0$, it tends to negative infinity. The latter case
will lead to no localized KK modes on the brane. Therefore, only the positive values of parameter $\lambda$ are considered
in the following. 

Furthermore, with $\lambda>0$, all solutions of KK modes obtained from Eq. (\ref{SchroAdS}) are bound states, including the
ground state. However, whether the ground state is the scalar zero mode or not is dependent on the shape of the effective
potential at finite area of extra dimension.

Moreover, in this case, at the origin of extra dimension, the effective potential (\ref{VzAdS}) becomes
\begin{eqnarray} \label{OriEff2AdS}
  V(z=0)=-2e^{(-1+e^{\frac{4H^2t_2(3\delta+2)}{\delta}})t_1}u^2+\frac{3H^2}{2\delta}
         +\frac{4H^4t_1t_2(\delta-1)(3\delta+2)}{\delta^3}e^{\frac{4H^2t_2(3\delta+2)}{\delta}}.
\end{eqnarray}
For this expression, if the parameter $u=0$, the effective potential will always be positive at position $z=0$. However,
if $u\neq0$, the first term within the expression of the r.h.s. of Eq. (\ref{OriEff2AdS}), which contains a 
double-exponential function, will predominantly determine the value of $V(z=0)$. Therefore, the negative $V(z=0)$ could 
exist, and the zero mode can serve as the ground state and be localized
on the brane through fine-tuning of parameters. To illustrate this, we fix the parameter $\lambda=1$ and choose an
appropriate $u$ to localize the zero mode. The shapes of the localized zero mode and the effective potential are depicted in
Fig. \ref{2IDWAdS} with parameters $H=1$, $\delta=2$, $\lambda=1$, $u=0.383$, $t_1=1$, and $t_2=0.1$. The mass spectra
of lower localized massive modes are also shown in the same figure. It can be observed that the zero mode is indeed the
ground state, and is localized on the brane along with all massive modes.
\begin{figure}[htb]
\begin{center}
\subfigure[$\tilde{\chi}_0(z)$]{\label{ZM2AdS}
\includegraphics[width= 0.45\textwidth]{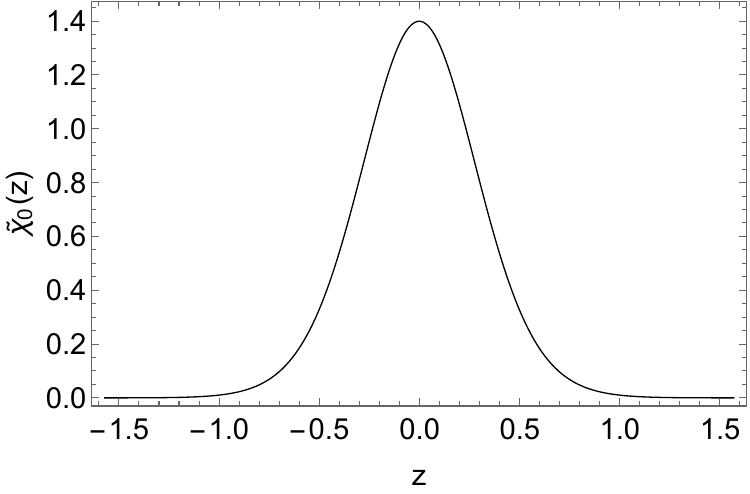}}
\subfigure[$V(z)$]{\label{Spec2IDWAdS}
\includegraphics[width= 0.45\textwidth]{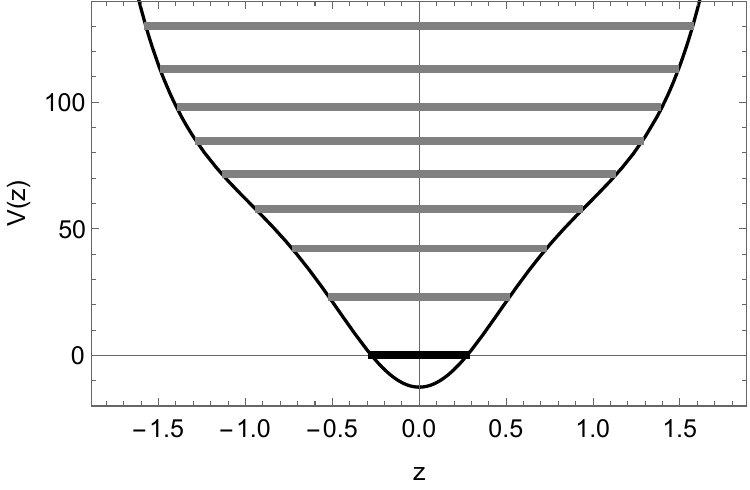}}
\end{center}\vskip -5mm
\caption{The scalar zero mode $\tilde{\chi}_0(z)$ and the effective potential $V(z)$ with parameters
         $H=1,\delta=2,\lambda=1,u=0.383,t_1=1,t_2=0.1$; the thick black line corresponds to $m_0^2=0$,
         and the first eight massive levels of the mass spectra are given by the grey lines.}
 \label{2IDWAdS}
\end{figure}

Therefore, in the AdS brane case, the zero mode can be localized on the brane by fine-tuning of parameters. The massive
KK modes can be localized on the thick brane, forming an infinite discrete spectra of mass. Additionally, if the bulk spacetime
is asymptotically flat, the coupling factor $F(R)\rightarrow1$, and has no influence at the boundaries of the extra dimension.

}

%%%%%%%%%%%%%%%%%%%%%%%%%%%%%%%%%%%%%%%%%
\section{Conclusions}\label{Cons}

In this paper, we consider a coupling mechanism between the kinetic term and the spacetime. There is a factor $F(R)$
which is a function of scalar curvature of the bulk in the five-dimensional action for scalars. Based on this scenario,
we explore the localization of the scalar field on specific braneworld models characterized by three different types of
brane geometries: Minkowski, dS$_4$, and AdS$_4$. The three brane models are regular with no singularity for the scalar
curvature, and are determined to be stable against the scalar perturbations.

In the case of a Minkowski brane, the scalar zero mode can always be localized on the brane. For the massive KK modes, there
is a critical value $\frac{1}{6}a^{2}$ for the coupling parameter $t$. For $t<\frac{1}{6}a^{2}$, the effective potential in
corresponding Schr\"{o}dinger-like equation for scalar KK mode tends to zero when far away from the brane, which leads to no
localized massive modes on the brane. However, with specific values of parameters, the resonant KK modes can exist in this case.
Then, for $t=\frac{1}{6}a^{2}$, there will be a P\"{o}schl-Teller-like effective potential, and a finite number of massive KK
modes can be localized on the brane. Lastly, for the case of $t>\frac{1}{6}a^{2}$, the effective potential exhibits an infinitely
deep well, and all the KK modes are trapped on the brane. Besides, when parameter $t<0$, the effective potential could be
positive at the origin of the extra dimension, giving rise to a local minimum shaping the scalar zero mode at the same position.

For the dS$_4$ brane case, we examine an asymptotically flat brane model, and suggest two forms for the
coupling factor $F(R)$. Upon introducing any one coupling factor, the scalar zero mode remains localized on the brane,
and the massive modes can be quasi-localized on the brane. Moreover, when considering the second
coupling function, a potential well with a positive lower boundary can emerge at position $z=0$. This
configuration gives rise to the localization of the zero mode on both sides of the origin, and the
quasi-localization of massive KK modes on the origin.

For the AdS$_4$ brane, we examine an asymptotically flat brane model as well. With consideration of the coupling factor $F(R)$ only,
the localization of scalar KK modes is similar to that of the minimal coupling case: The scalar zero mode is non-normalized, and
cannot be localized; For the massive modes, an infinitely deep well exists, and all the massive modes are confined on the brane.
Then, if we introduce the coupling potential $V(\Phi,\varphi)$, the scalar zero mode can be localized on the brane by fine-tuning
of parameters, and there are infinite number of localized massive modes on the brane still.

Regarding the coupling factor $F(R)$, if the brane model is (asymptotically) flat, the factor $F(R)$ tends to the value $1$ and
has no influence at the boundaries of the extra dimension. Furthermore, for the localization of the scalar field in the minimal
coupling case, if the scalar zero mode cannot be localized, introducing the factor $F(R)$ will not lead to the localization of
this mode. Conversely, if the scalar zero mode is localized on the brane, the presence of the factor $F(R)$ can offer numerous
perspectives for this mode. {Besides, if the 5D spacetime is asymptotically AdS$_5$, the function $F(R)$ will
tend to a nonnegative constant (and not approach $1$) at the boundaries of extra dimension, and then the effective potential
of KK modes could behave diversely, and the KK modes can be endowed with more abundant information.}

\acknowledgments

The authors are extremely grateful for the anonymous
referees, whose comments led to the improvement of this paper.
This work is supported by the National Natural
Science Foundation of China (Grants No. 11305119), the Natural Science Basic Research Plan in Shaanxi Province of China
(Program No. 2020JM-198), the Fundamental Research Funds for the Central Universities (Grants No. JB170502),
and the 111 Project (B17035).

\end{document}